\documentclass[showpacs,twocolumn,nofootinbib]{revtex4}
\usepackage{eurosym}
\usepackage{amsfonts}
\usepackage{amsmath}
\usepackage{graphicx}
\usepackage{amssymb}
\usepackage{subfig}
\usepackage{xcolor}

\begin{document}

\title{Secular precessing compact binary dynamics, spin and orbital angular
momentum flip-flops}
\author{M\'{a}rton T\'{a}pai$^{\dag }$, Zolt\'{a}n Keresztes$^{\ddag }$, L%
\'{a}szl\'{o} \'{A}rp\'{a}d Gergely$^{\star }$}
\affiliation{Institute of Physics, University of Szeged, D\'{o}m t\'{e}r 9, Szeged 6720,
Hungary\\
$^{\dag }${\small E-mail: tapai@titan.physx.u-szeged.hu\quad }$^{\ddag }$%
{\small E-mail: zkeresztes@titan.physx.u-szeged.hu\quad}$^{\star }${\small \
E-mail: gergely@physx.u-szeged.hu }}

\begin{abstract}
We derive the conservative secular evolution of precessing compact binaries
to second post-Newtonian order accuracy, with leading-order spin-orbit,
spin-spin and mass quadrupole-monopole contributions included. The emerging
closed system of first-order differential equations evolves the pairs of
polar and azimuthal angles of the spin and orbital angular momentum vectors
together with the periastron angle. In contrast with the instantaneous
dynamics, the secular dynamics is autonomous. This secular dynamics reliably
characterizes the system over timescales starting from a few times the
radial period to several precessional periods, but less than the radiation
reaction timescale. We numerically compare the instantaneous and secular
evolutions and estimate the number of periods for which dissipation has no
significant effect, e.g. the conservative timescale. We apply the analytic
equations to study the spin flip-flop effect, recently found by numerical
relativity methods. Our investigations show that the effect does not
generalize beyond its original parameter settings, although we reveal
distinct configurations exhibiting one half flip-flops. In addition, we find
a flip-flopping evolution of the orbital angular momentum vector, which
ventures from one pole to another through several precessional periods. This
is a new effect, occurring for mass ratios much less than one.
\end{abstract}

\maketitle

\section{Introduction}

The LIGO collaboration with the improved Advanced LIGO detectors \cite{aLIGO}%
, has recently announced its first two detections of gravitational waves
from coalescing stellar mass black hole binaries \cite{detection1},\cite%
{detection2}. With the approaching second observation run, the era of
gravitational wave astrophysics has commenced.

Whenever the black hole spins $\mathbf{S}_{\mathbf{1}}$, $\mathbf{S}_{%
\mathbf{2}}$ and the orbital angular momentum $\mathbf{L}_{\mathbf{N}}$ of
the binary do not align, they undergo precession. The total angular momentum 
$\mathbf{J}=\mathbf{L}_{\mathbf{N}}+\mathbf{S}_{\mathbf{1}}+\mathbf{S}_{%
\mathbf{2}}$ is conserved up to the 2 post-Newtonian (PN) order. The
gravitational waveform emitted by the system exhibits both amplitude and
phase modulations due to precession. Detection of precessing gravitational
waves requires corresponding waveforms, such as the analytic frequency
domain waveform developed in Ref. \cite{CKCY}. The consequences of
precession on gravitational waveforms were presented in Refs. \cite{KCY}-%
\cite{PhenomPwf}. In Refs. \cite{GuptaGopakumar}-\cite{OBKPSSZ} the
precessing conventions and various spin related contributions were
discussed, while Refs. \cite{HPBB}-\cite{OFOCRKL} analysed the searches for
spin-precessing binaries. The 4PN spin dependent conservative dynamics of
inspiralling compact binaries was completed in Ref. \cite{LS4PN}. A fully
precessing analysis of the first gravitational wave detection GW150914 was
presented in Ref. \cite{precGW150914}.

Compact binary dynamics in the inspiral regime exhibit three distinct
timescales. The shortest is the radial timescale, defined by the radial
period. The precessional timescale is given by the time necessary for the
orbital angular momentum $\mathbf{L}_{\mathbf{N}}$ or for the spins $\mathbf{%
S}_{\mathbf{1}}$, $\mathbf{S}_{\mathbf{2}}$ to undergo a full rotation about
their precession axis. The third is the gravitational radiation reaction
timescale, over which the effects of gravitational dissipation are
noticeable.

Averaging the dynamics over some of these timescales may turn useful. When
precession related effects are examined, averaging over a radial period will
remove insignificant instantaneous effects, but keeps the precessional
evolution, which will be dominant. Precession related effects like the
spin-flip \cite{spinflip}, explaining the formation of X-shape radio
galaxies \cite{xshaped}, and transitional precession \cite{ACST} were
examined with this method. Precession of the dominant spin of a supermassive
black hole binary was also identified for the first time from VLBI radio
data of its jet spanning over 18 years in Ref. \cite{VLBIKun}. For
gravitational dissipation, averaging over the precessional period may turn
useful.

The instantaneous dynamics (including spin effects) in terms of
dimensionless variables was discussed in Refs. \cite{Inspiral1} - \cite%
{chameleon}. This generalizes earlier works on binary dynamics of Refs. \cite%
{damourderuelle} - \cite{KMG}.

In this paper we investigate compact binary systems which are subject to
bound motion, establishing the 2PN secular dynamics. In Section \ref{timeeq}
we will derive the radial period in terms of the averaged values of the
dimensionless orbital angular momentum 
\begin{equation}
\mathfrak{l}_{r}=\frac{cL_{N}}{Gm\mu }~,
\end{equation}%
and dimensionless eccentricity%
\begin{equation}
e_{r}=\frac{A_{N}}{Gm\mu }~.
\end{equation}%
These two parameters characterize the shape of the orbit in the plane of the
motion. Hence they will be dubbed shape variables. The total mass of the
binary is $m=m_{1}+m_{2}$, $\mu =m_{1}m_{2}/m$ is the reduced mass, $G$ is
the gravitational constant, $c$ is the speed of light and $A_{N}$ is the
magnitude of the Laplace-Runge-Lenz vector. Other notations frequently used
in this paper are the mass ratio $\nu =m_{2}/m_{1}$ and the symmetric mass
ratio $\eta =\mu /m$. The magnitude of the spins is given by the
dimensionless spin parameters $\chi _{i}$ $\left( i=1,2\right) $. A
dimensionless time variable $\mathfrak{t}=tc^{3}/Gm$ (with time $t$) was
introduced in Ref. \cite{chameleon}. Derivatives with regards to the
dimensionless time are denoted by a dot.

Auxiliary calculations required for
the radial period will be given in Appendices \ref{appendix1}-\ref{inverting}%
. Appendix \ref{appendix1} presents the radial period in terms of the
variables evaluated at the periastron (characterized by the value of the
true anomaly parameter $\chi _{p}=0$). The $\chi _{p}$ dependence of the
shape variables is also derived in this Appendix. Appendix \ref{lreraver}
contains the averaged shape variables. In Appendix \ref{inverting} we
express the shape variables evaluated in the periastron with the
corresponding averaged quantities.

In Section \ref{secdyn} we present the main result of the paper, which is
the \textit{secular} precessing compact binary dynamics. For completeness,
in Appendix \ref{appendix2} we give the secular precession angular
velocities. In Section \ref{limits} we investigate the limits of validity of
the secular dynamics, establishing the corresponding conservative timescale.

When the dominant spin vector approximately lies in the plane of motion,
while the smaller spin is closely aligned with the orbital angular momentum $%
\mathbf{L}_{N}$, the smaller spin slowly evolves to be anti aligned with $%
\mathbf{L}_{N}$, then periodically changes back and forth on a timescale
shorter than the gravitational radiation reaction timescale. This effect,
dubbed flip-flop has been established by numerical relativity methods in
Refs. \cite{flipflop} - \cite{flipflop2}. In Section \ref{spinflipflop} we
examine the parameter dependence of the spin flip-flop phenomenon, as an
application of the derived secular dynamics. We reproduce analytically the
flip-flop effect in its original parameter settings, then we find parameter
ranges where only a half flip-flop occurs, but for both spins.

In Section \ref{lnflipflop} we discuss a parameter configuration for which
the orbital angular momentum undergoes a similar flip-flopping, a behaviour
unaccounted before.

In Section \ref{concludingr} we give the conclusions.

\section{Integration of the instantaneous dynamics \label{timeeq}}

The 2PN conservative dynamics of compact binary systems was given as Eqs.
(36)-(42) of Ref. \cite{chameleon} in terms of dimensionless osculating
orbital elements $\mathfrak{l}_{r}$, $e_{r}$, $\psi _{p}$, $\alpha $, $\phi
_{n}$, spin polar and azimuthal angles $\kappa _{i}$ and $\zeta _{i}$ ($%
i=1,2 $), the true anomaly parametrization $\chi _{p}$. The time evolution
of $\chi _{p}$ is governed by Eq. (43) of Ref. \cite{chameleon}. The polar
and azimuthal angles of the spins are defined in a system with $\mathbf{L}%
_{N}$ as the z-axis and $\mathbf{A}_{N}$ as the x-axis. The argument of the
periastron, $\psi _{p}$ is defined by $\psi _{p}=\arccos \left( \mathbf{\hat{%
l}}\cdot \mathbf{\hat{A}}_{N}\right) $, with $\mathbf{\hat{l}}=\mathbf{\hat{J%
}}\times \mathbf{\hat{L}}_{N}$. The inclination $\alpha $ is the polar angle
of $\mathbf{\hat{L}}_{N}$ in the system where the z-axis points in the
direction of $\mathbf{\hat{J}}$, and the x-axis is given by $\mathbf{\hat{l}}
$. The last angle is the longitude of the ascending node $-\phi _{n}$, span
by the inertial axis $\mathbf{\hat{x}}$ (arbitrarily chosen in the plane
perpendicular to $\mathbf{\hat{J}}$) and $\mathbf{\hat{l}}$. This angle
becomes the azimuthal angle of $\mathbf{\hat{L}}_{N}$, if initially $\mathbf{%
\hat{L}}_{N}$ is set in the plane defined by $\mathbf{\hat{x}}$ and $\mathbf{%
\hat{J}}$.

The main purpose of this paper is to average this instantaneous dynamics
over a radial period, to 2PN order. For bounded motion, by definition the
dimensionless period $\mathfrak{T}\equiv Tc^{3}/Gm$ , referring to a change
in the true anomaly $\chi _{p}\in \left\{ 0,2\pi \right\} $, can be computed
as 
\begin{equation}
\mathfrak{T}\equiv \int_{0}^{\mathfrak{T}}d\mathfrak{t}=\int_{0}^{2\pi }%
\frac{1}{\dot{\chi}_{p}}d\chi _{p}~.  \label{perioddef}
\end{equation}%
By formally integrating (\ref{perioddef}) we get the following PN expansion:%
\begin{eqnarray}
\mathfrak{T} &=&\mathfrak{T}_{0}\left( 1+\frac{\tau _{0PN}}{\mathfrak{l}%
_{r}^{2}}+\frac{\tau _{0SO}}{\mathfrak{l}_{r}^{3}}+\frac{\tau _{0SS}}{%
\mathfrak{l}_{r}^{4}}\right.  \notag \\
&&\left. +\frac{\tau _{0QM}}{\mathfrak{l}_{r}^{4}}+\frac{\tau _{02PN}}{%
\mathfrak{l}_{r}^{4}}\right) ~.  \label{period}
\end{eqnarray}%
Here $1/\mathfrak{l}_{r}^{2}$ stands for one PN\ order, as explained in Ref. 
\cite{chameleon}, while the lower index $0$ indicates values taken at $\chi
_{p}=0$. In order to explicitly compute the terms of Eq. (\ref{period}), the 
$\chi _{p}$ dependence of $\mathfrak{l}_{r}$ and $e_{r}$ are required. Their
derivation and explicit expressions are given in Appendix \ref{appendix1}.

A different expansion of the period arises in terms of the averaged shape
variables $\mathfrak{\bar{l}}_{r}$\ and $\bar{e}_{r}$:%
\begin{eqnarray}
\mathfrak{T} &=&\mathfrak{\tilde{T}}\left( 1+\frac{1}{\mathfrak{\bar{l}}%
_{r}^{2}}\tilde{\tau}_{PN}+\frac{1}{\mathfrak{\bar{l}}_{r}^{3}}\tilde{\tau}%
_{SO}+\frac{1}{\mathfrak{\bar{l}}_{r}^{4}}\tilde{\tau}_{QM}\right.  \notag \\
&&\left. +\frac{1}{\mathfrak{\bar{l}}_{r}^{4}}\tilde{\tau}_{SS}+\frac{1}{%
\mathfrak{\bar{l}}_{r}^{4}}\tilde{\tau}_{2PN}\right) ~.  \label{periodaver}
\end{eqnarray}%
Here%
\begin{eqnarray}
\mathfrak{\tilde{T}} &=&\frac{2\mathfrak{\bar{l}}_{r}^{3}\pi }{\left( 1-\bar{%
e}_{r}^{2}\right) ^{3/2}}~, \\
\tilde{\tau}_{PN} &=&\sqrt{1-\bar{e}_{r}^{2}}(15-9\eta )  \notag \\
&&+\left( 1-\bar{e}_{r}^{2}\right) (7\eta -6)~, \\
\tilde{\tau}_{SO} &=&0~, \\
\tilde{\tau}_{2PN} &=&\frac{\sqrt{1-\bar{e}_{r}^{2}}}{64\bar{e}_{r}^{4}}%
\sum_{k=0}^{6}\bar{U}_{k}\bar{e}_{r}^{k}  \notag \\
&&-\frac{(\bar{e}_{r}+1)}{8\bar{e}_{r}^{4}}\sum_{k=0}^{7}\bar{V}_{k}\bar{e}%
_{r}^{k}~,  \label{tau2pnkif} \\
\tilde{\tau}_{QM} &=&\frac{3\eta }{512\bar{e}_{r}\left( 1-\bar{e}%
_{r}^{2}\right) ^{2}\mathfrak{\bar{l}}_{r}^{4}}\sum_{k=1}^{2}\chi
_{k}^{2}\nu ^{2k-3}w_{k}  \notag \\
&&\times \left[ \bar{U}^{QM}\sin ^{2}\kappa _{k}\cos 2\zeta _{k}\right. 
\notag \\
&&\left. +\bar{V}^{QM}\left( 3\cos 2\kappa _{k}\allowbreak +1\right) \right]
~, \\
\bar{\tau}_{SS} &=&\frac{3\chi _{1}\chi _{2}\eta }{8(1-\bar{e}_{r})^{2}\bar{e%
}_{r}}\left( \cos \kappa _{1}\cos \kappa _{2}\bar{U}^{SS}\right.  \notag \\
&&\left. +\sin \kappa _{1}\sin \kappa _{2}\bar{V}^{SS}\right) ~.
\label{tausskif}
\end{eqnarray}%
$\allowbreak $The coefficients in the above expressions are enlisted in
Table \ref{tabletauaver2pn}. 
\begin{table}[tbp]
\caption{The coefficients in Eqs (\protect\ref{tau2pnkif})-(\protect\ref%
{tausskif}).}
\label{tabletauaver2pn}
\begin{center}
\begin{tabular}{cc}
\hline\hline
$Coefficient$ & $Expression$ \\ \hline\hline
$\bar{U}_{6}$ & $-437\eta ^{2}+3336\eta -1008$ \\ \hline
$\bar{U}_{5}$ & $-64\left( 8\eta ^{2}-6\eta -5\right) $ \\ \hline
$\bar{U}_{4}$ & $-8\left( 211\eta ^{2}-159\eta +336\right) $ \\ \hline
$\bar{U}_{3}$ & $64\left( 4\eta ^{2}+11\eta -5\right)$ \\ \hline
$\bar{U}_{2}$ & $-8\left( 79\eta ^{2}-600\eta +528\right)$ \\ \hline
$\bar{U}_{1}$ & $-128\left( \eta ^{2}-8\eta +15\right)$ \\ \hline
$\bar{U}_{0}$ & $32\left( 65\eta ^{2}-238\eta +180\right)$ \\ \hline\hline
$\bar{V}_{7}$ & $224\eta ^{2}-690\eta +360$ \\ \hline
$\bar{V}_{6}$ & $2\left( 64\eta ^{2}-11\eta -12\right) $ \\ \hline
$\bar{V}_{5}$ & $139\eta ^{2}-410\eta +452$ \\ \hline
$\bar{V}_{4}$ & $-179\eta ^{2}+266\eta -308$ \\ \hline
$\bar{V}_{3}$ & $-27\eta ^{2}+28\eta +8$ \\ \hline
$\bar{V}_{2}$ & $67\eta ^{2}-4\eta +72 $ \\ \hline
$\bar{V}_{1}$ & $-12\left( 23\eta ^{2}-90\eta +80\right) $ \\ \hline
$\bar{V}_{0}$ & $260\eta ^{2}-952\eta +720$ \\ \hline\hline
$\bar{U}^{QM}$ & $-4\left( 27\bar{e}_{r}^{7}-72\bar{e}_{r}^{6}+263\bar{e}%
_{r}^{5}\right.$ \\ 
& $-1674\bar{e}_{r}^{4}-1702\bar{e}_{r}^{3}-4116\bar{e}_{r}^{2}$ \\ 
& $\left. -1360\bar{e}_{r}-960\right)$ \\ \hline
$\bar{V}^{QM}$ & $16\left( \bar{e}_{r}^{7}-2\bar{e}_{r}^{6}+9\bar{e}%
_{r}^{5}-43\bar{e}_{r}^{4}\right.$ \\ 
& $\left. -69\bar{e}_{r}^{3}-108\bar{e}_{r}^{2}-46\bar{e}_{r}-12\right)$ \\ 
\hline\hline
$\bar{U}^{SS}$ & $-8\bar{e}_{r}^{5}+21\bar{e}_{r}^{4}-15\bar{e}_{r}^{3}$ \\ 
& $-38\bar{e}_{r}^{2}-56\bar{e}_{r}-24$ \\ \hline
$\bar{V}^{SS}$ & $\frac{\cos (\zeta _{1}-\zeta _{2})}{2}\allowbreak \left( 8%
\bar{e}_{r}^{5}-21\bar{e}_{r}^{4}\right.$ \\ 
& $\left. +15\bar{e}_{r}^{3}+38\bar{e}_{r}^{2}+56\bar{e}_{r}+24\right)$ \\ 
& $+\frac{\cos (\zeta _{1}+\zeta _{2})}{4\bar{e}_{r}^{2}(\bar{e}_{r}+1)}%
\left[ 371\bar{e}_{r}^{7}-276\bar{e}_{r}^{6}\right.$ \\ 
& $+\bar{e}_{r}^{5}\left( -48\bar{e}_{r}+104\sqrt{1-\bar{e}_{r}^{2}}%
+1771\right)$ \\ 
& $-\bar{e}_{r}^{4}\left( 48\bar{e}_{r}+104\sqrt{1-\bar{e}_{r}^{2}}%
+1517\right)$ \\ 
& $+8\bar{e}_{r}^{3}\left( 24\bar{e}_{r}+583\sqrt{1-\bar{e}_{r}^{2}}%
-854\right)$ \\ 
& $+4\bar{e}_{r}^{2}\left( 48\bar{e}_{r}-1166\sqrt{-\bar{e}_{r}^{2}+1}%
+1881\right)$ \\ 
& $-8\left( 596\sqrt{1-\bar{e}_{r}^{2}}-593\right) \bar{e}_{r}$ \\ 
& $\left. +4768\left( \sqrt{1-\bar{e}_{r}^{2}}-1\right) \right]$ \\ 
\hline\hline
\end{tabular}%
\end{center}
\end{table}

The time average $\bar{f}$ of any quantity $f\left( \mathfrak{t}\right) $
with respect to $\mathfrak{t}$ has been defined as%
\begin{equation}
\mathfrak{T}\bar{f}=\int_{0}^{\mathfrak{T}}f\left( \mathfrak{t}\right) d%
\mathfrak{t}=\int_{0}^{2\pi }f\left( \chi _{p}\right) \frac{1}{\dot{\chi}_{p}%
}d\chi _{p}~.  \label{averdef}
\end{equation}%
We note that for the change of variables from $\mathfrak{t}$ to $\chi _{p}$
to be valid at a certain PN accuracy, $\mathfrak{T}$ needs to be given at
the same PN accuracy as $f\left( \chi _{p}\right) .$

Hence the average expressions of $\mathfrak{l}_{r}$ and $e_{r}$ are computed
as

\begin{eqnarray}
\mathfrak{\bar{l}}_{r} &=&\frac{1}{\mathfrak{T}}\int_{0}^{2\pi }\frac{%
\mathfrak{l}_{r}\left( \chi _{p}\right) }{\dot{\chi}_{p}}d\chi _{p}~,
\label{lraverdefkif} \\
\bar{e}_{r} &=&\frac{1}{\mathfrak{T}}\int_{0}^{2\pi }\frac{e_{r}\left( \chi
_{p}\right) }{\dot{\chi}_{p}}d\chi _{p}~.  \label{eraverdefkif}
\end{eqnarray}%
Integrating and Taylor-expanding to 2PN order accuracy leads to the
following formal expressions%
\begin{eqnarray}
\mathfrak{\bar{l}}_{r} &\!=\!&\frac{\mathfrak{l}_{r0N}\!+\!\mathfrak{\bar{l}}%
_{rPN}\!+\!\mathfrak{\bar{l}}_{rSO}\!+\!\mathfrak{\bar{l}}_{rSS}\!+\!%
\mathfrak{\bar{l}}_{rQM}\!+\!\mathfrak{\bar{l}}_{r2PN}}{\mathfrak{T}}~,
\label{lraverexpr} \\
\bar{e}_{r} &\!=\!&\frac{e_{r0N}\!+\!\bar{e}_{rPN}\!+\!\bar{e}_{rSO}\!+\!%
\bar{e}_{rSS}\!+\!\bar{e}_{rQM}\!+\!\bar{e}_{r2PN}}{\mathfrak{T}}~.
\label{eraverexpr}
\end{eqnarray}%
(The radial period is taken at 2PN accuracy.) The computation leading to the
averaged shape variables is given in Appendix \ref{lreraver}, while the
derivation of the shape variables at $\chi _{p}=0$ in terms of the averaged
quantities $\mathfrak{\bar{l}}_{r}$\ and $\bar{e}_{r}$ in Appendix \ref%
{inverting}. \captionsetup[subfigure]{position=top} 
\begin{figure*}[tbp]
\subfloat[\large{\textcolor{green}{\line(1,0){15}} Secular, \textcolor{red}{\line(1,0){15}} Instantaneous\label{subfig-1a}}]{$
\begin{array}{ccc}
\includegraphics[scale=0.6]{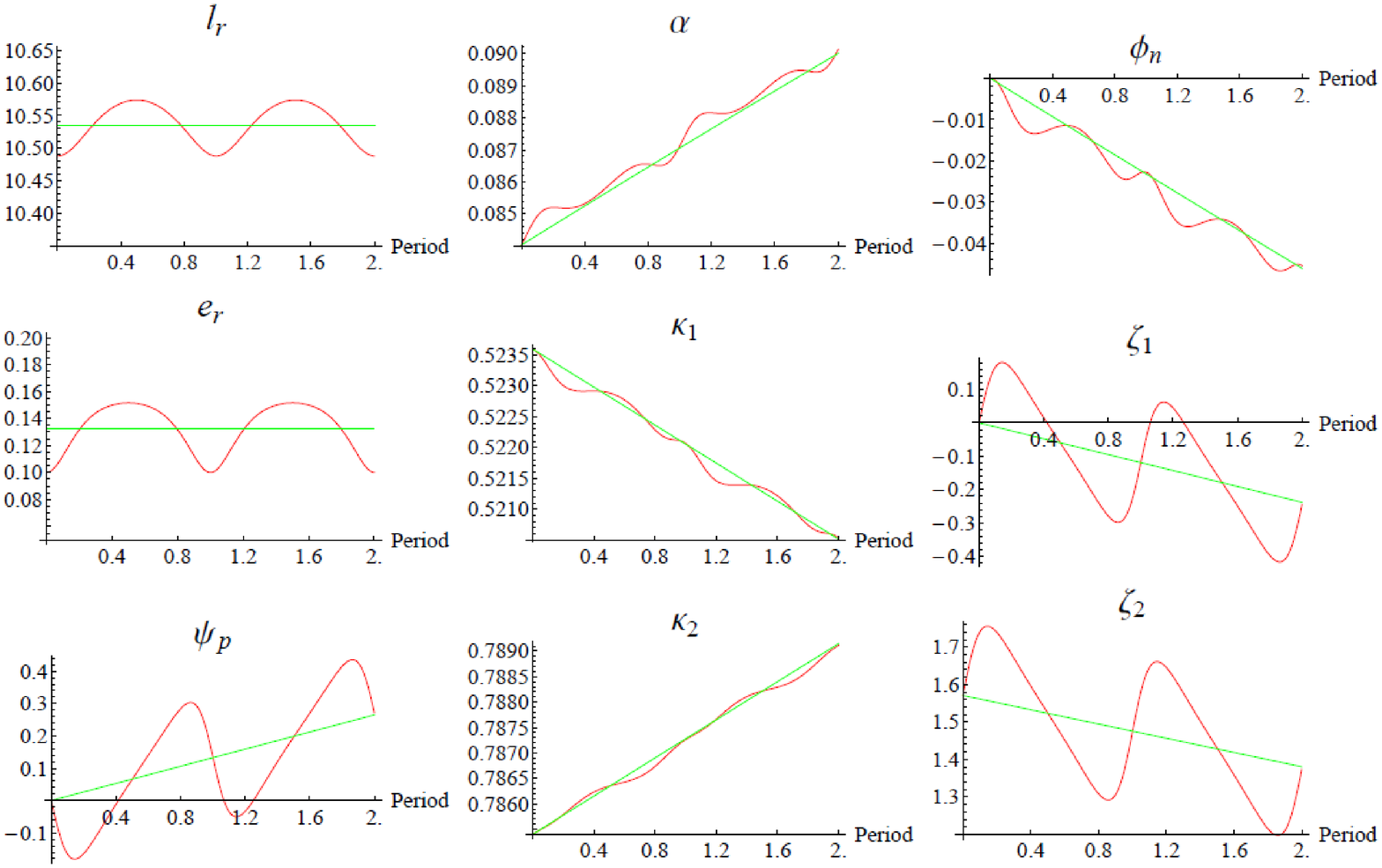}
\end{array}
$

    } \vfill
\subfloat[\large{\textcolor{green}{\line(1,0){15}} Secular, \textcolor{red}{\line(1,0){15}} Instantaneous\label{subfig-1b}}]{$
\begin{array}{ccc}
\includegraphics[scale=0.6]{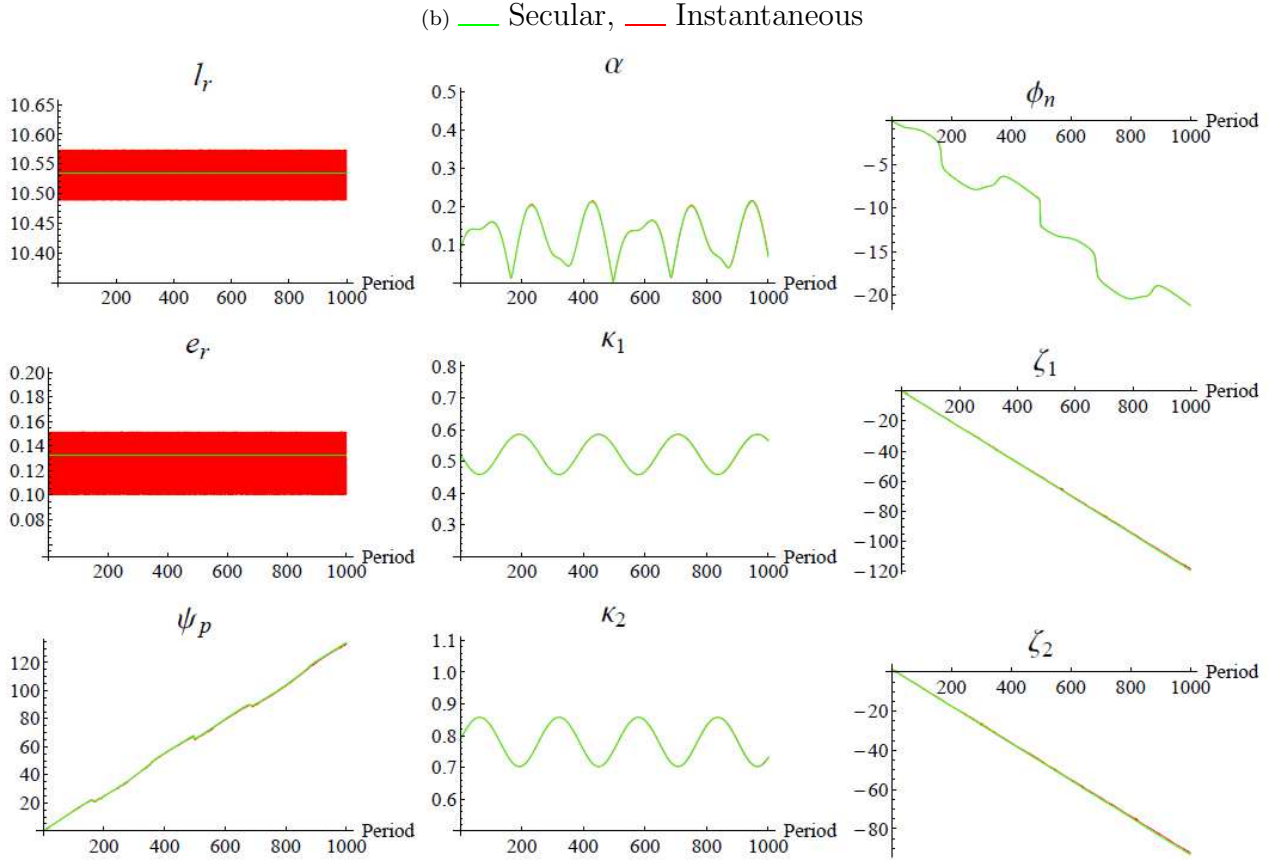}
\end{array}
$
    }
\caption{Comparison of the secular and instantaneous dynamics. The upper
block a) shows the first two periods, while the lower block b) is over the
conservative timescale (shown as the corresponding number of periods on the
x-axis). The figures are for total mass $m=20M_{\odot }$, mass ratio $%
\protect\nu =0.5$, dimensionless spin parameters $\protect\chi _{1}=0.9982$, 
$\protect\chi _{2}=0.9982$, PN parameter $\protect\varepsilon =0.01$. The
initial values of the other parameters are $e_{r}=0.1$, $\protect\kappa _{1}=%
\protect\pi /6$, $\protect\kappa _{2}=\protect\pi /4$, $\protect\zeta _{1}=0$%
, $\protect\zeta _{2}=\protect\pi /2$, $\protect\psi _{p}=0$, $\protect\phi %
_{n}=0$.}
\label{compfig1}
\end{figure*}

\begin{figure*}[tbp]
\subfloat[\large{\textcolor{green}{\line(1,0){15}} Secular, \textcolor{red}{\line(1,0){15}} Instantaneous\label{subfig-2a}}]{$
\begin{array}{ccc}
\includegraphics[scale=0.6]{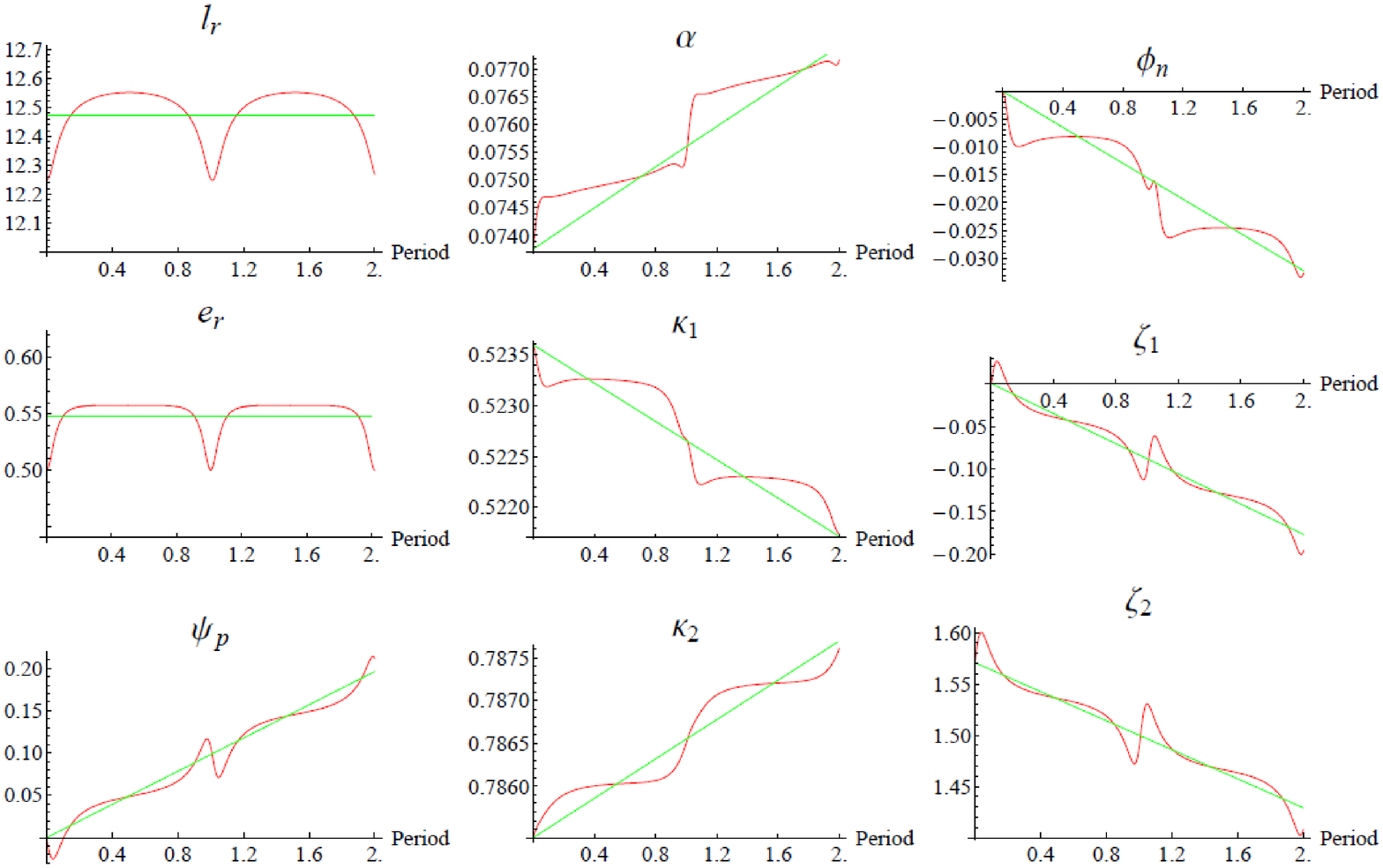} 
\end{array}
$

    } \vfill
\subfloat[\large{\textcolor{green}{\line(1,0){15}} Secular, \textcolor{red}{\line(1,0){15}} Instantaneous\label{subfig-2b}}]{$
\begin{array}{ccc}
\includegraphics[scale=0.6]{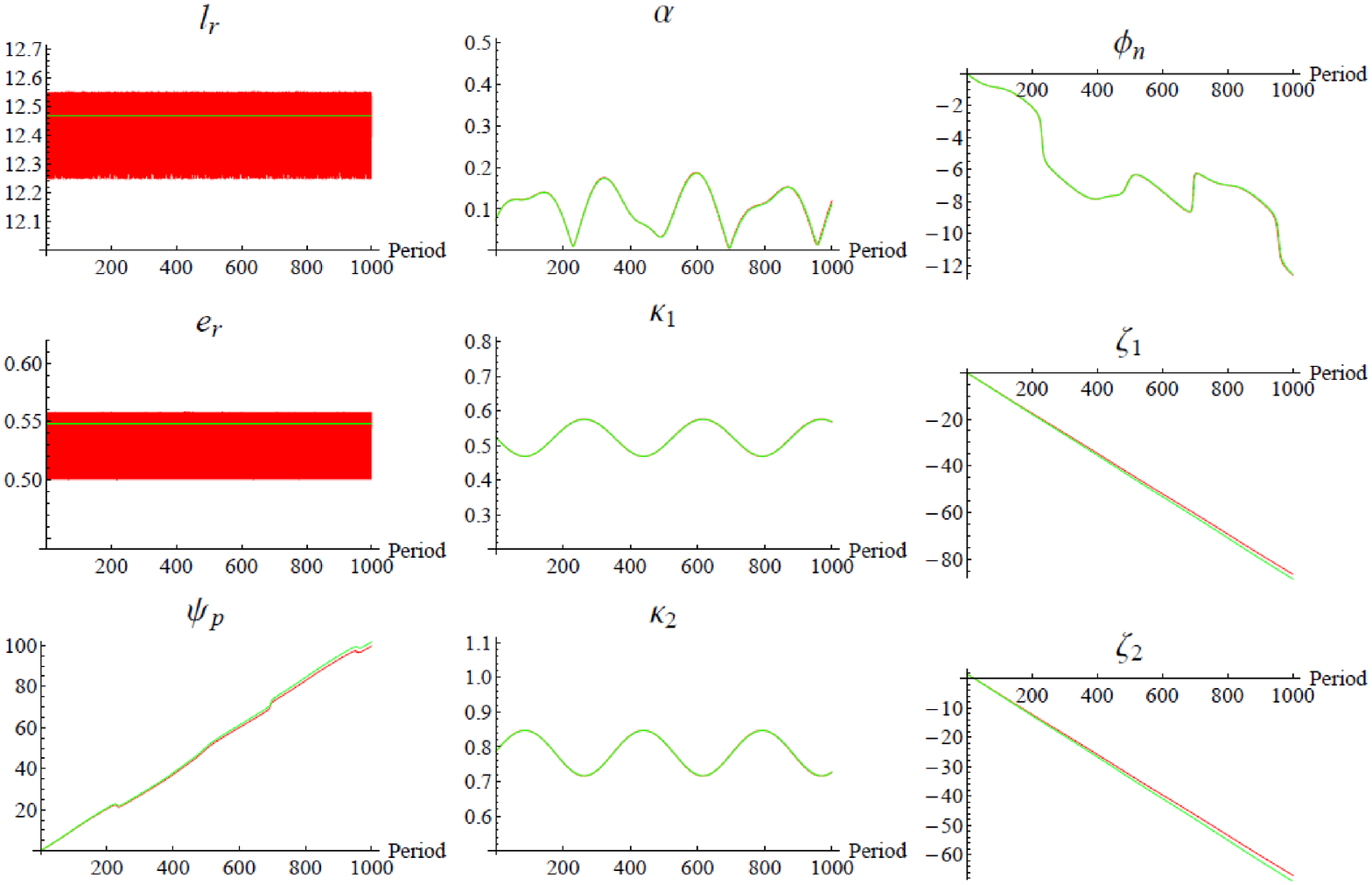}
\end{array}
$
    }
\caption{Comparison of the secular and instantaneous dynamics. The upper
block a) shows the first two periods, while the lower block b) is over the
conservative timescale (shown as the corresponding number of periods on the
x-axis). The figures are for total mass $m = 20 M_{\odot}$, mass ratio $%
\protect\nu = 0.5$, dimensionless spin parameters $\protect\chi_{1} = 0.9982$%
, $\protect\chi_{2} = 0.9982$, PN parameter $\protect\varepsilon = 0.01$.
The initial values of the other parameters are $e_{r}=0.5$, $\protect\kappa%
_{1}=\protect\pi/6$, $\protect\kappa_{2}=\protect\pi/4$, $\protect\zeta%
_{1}=0 $, $\protect\zeta_{2}=\protect\pi/2$, $\protect\psi_{p}=0$, $\protect%
\phi_{n}=0$.}
\label{compfig2}
\end{figure*}

\begin{figure*}[tbp]
\subfloat[\large{\textcolor{green}{\line(1,0){15}} Secular, \textcolor{red}{\line(1,0){15}} Instantaneous\label{subfig-3a}}]{$
\begin{array}{ccc}
\includegraphics[scale=0.6]{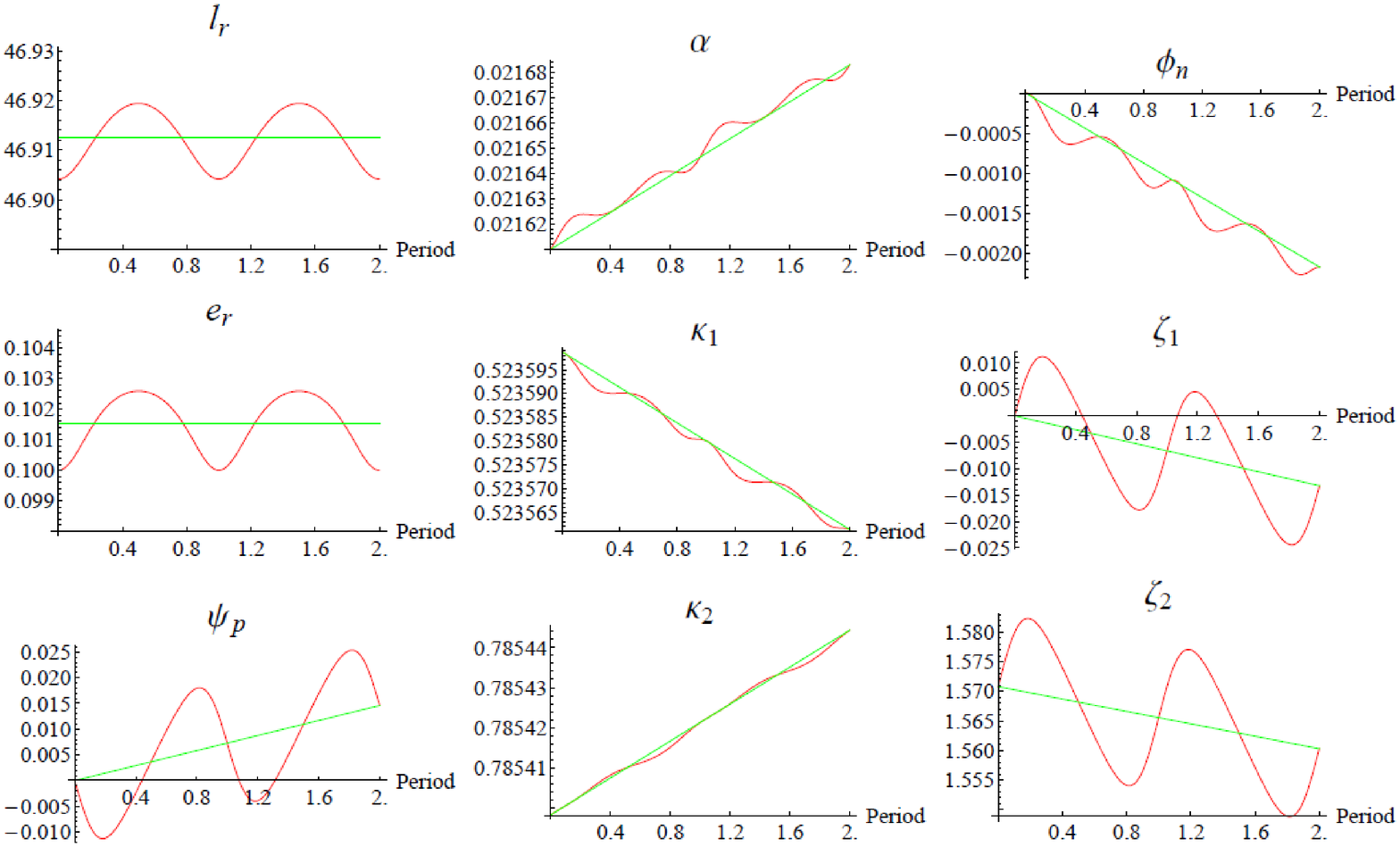} 
\end{array}
$

    } \vfill
\subfloat[\large{\textcolor{green}{\line(1,0){15}} Secular, \textcolor{red}{\line(1,0){15}} Instantaneous\label{subfig-3b}}]{$
\begin{array}{ccc}
\includegraphics[scale=0.6]{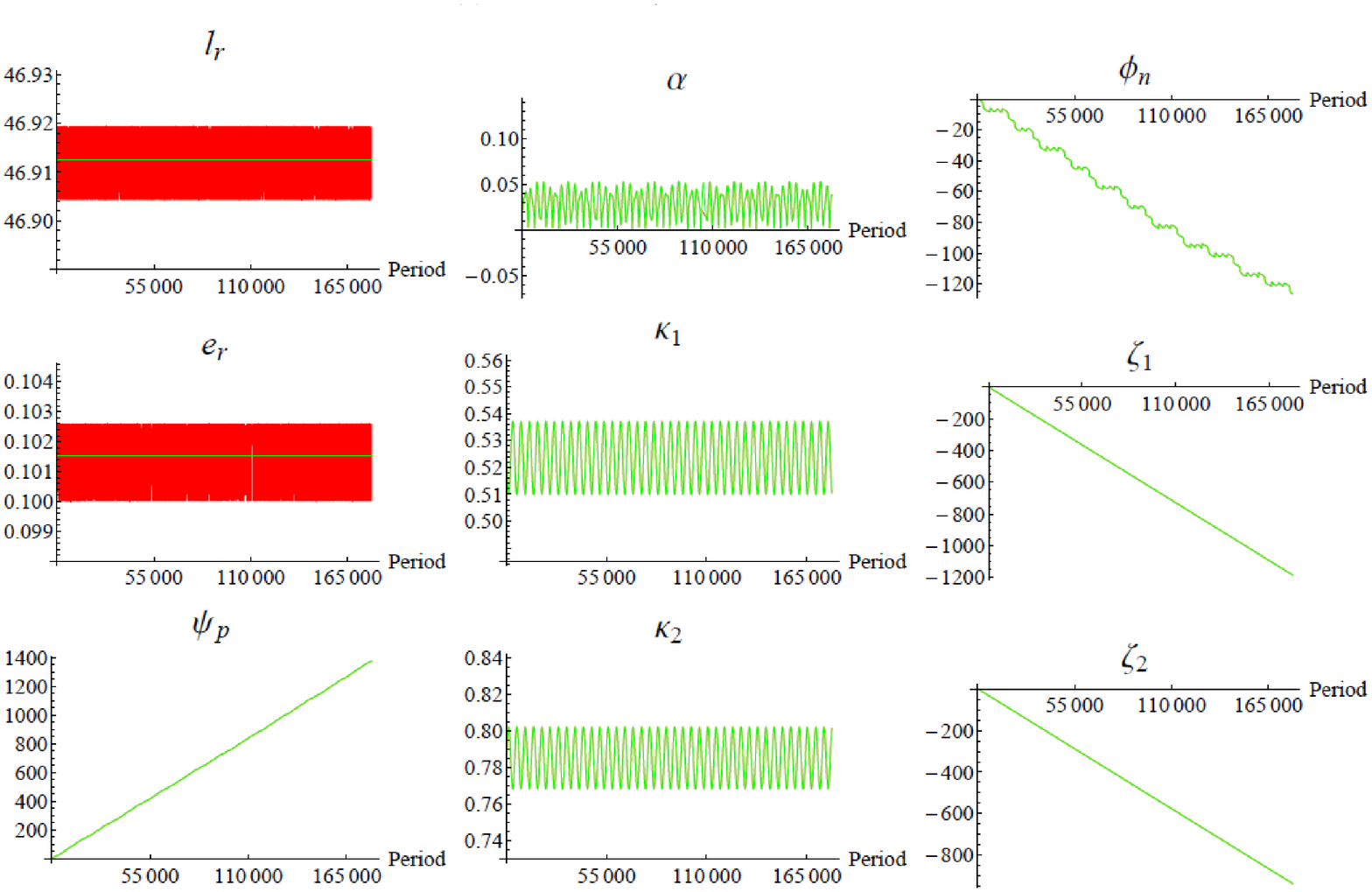}
\end{array}
$
    }
\caption{Comparison of the secular and instantaneous dynamics. The upper
block a) shows the first two periods, while the lower block b) is over the
conservative timescale (shown as the corresponding number of periods on the
x-axis, the conservative timescale being set as $N=0.001\protect\varepsilon%
^{-5/2}$ number of periods). The figures are for total mass $m = 20
M_{\odot} $, mass ratio $\protect\nu = 0.5$, dimensionless spin parameters $%
\protect\chi_{1} = 0.9982$, $\protect\chi_{2} = 0.9982$, PN parameter $%
\protect\varepsilon = 0.0005$. The initial values of the other parameters
are $e_{r}=0.1$, $\protect\kappa_{1}=\protect\pi/6$, $\protect\kappa_{2}=%
\protect\pi/4$, $\protect\zeta_{1}=0 $, $\protect\zeta_{2}=\protect\pi/2$, $%
\protect\psi_{p}=0$, $\protect\phi_{n}=0$.}
\label{compfig3}
\end{figure*}

\begin{figure*}[tbp]
\subfloat[\large{\textcolor{green}{\line(1,0){15}} Secular, \textcolor{red}{\line(1,0){15}} Instantaneous\label{subfig-4a}}]{$
\begin{array}{ccc}
\includegraphics[scale=0.6]{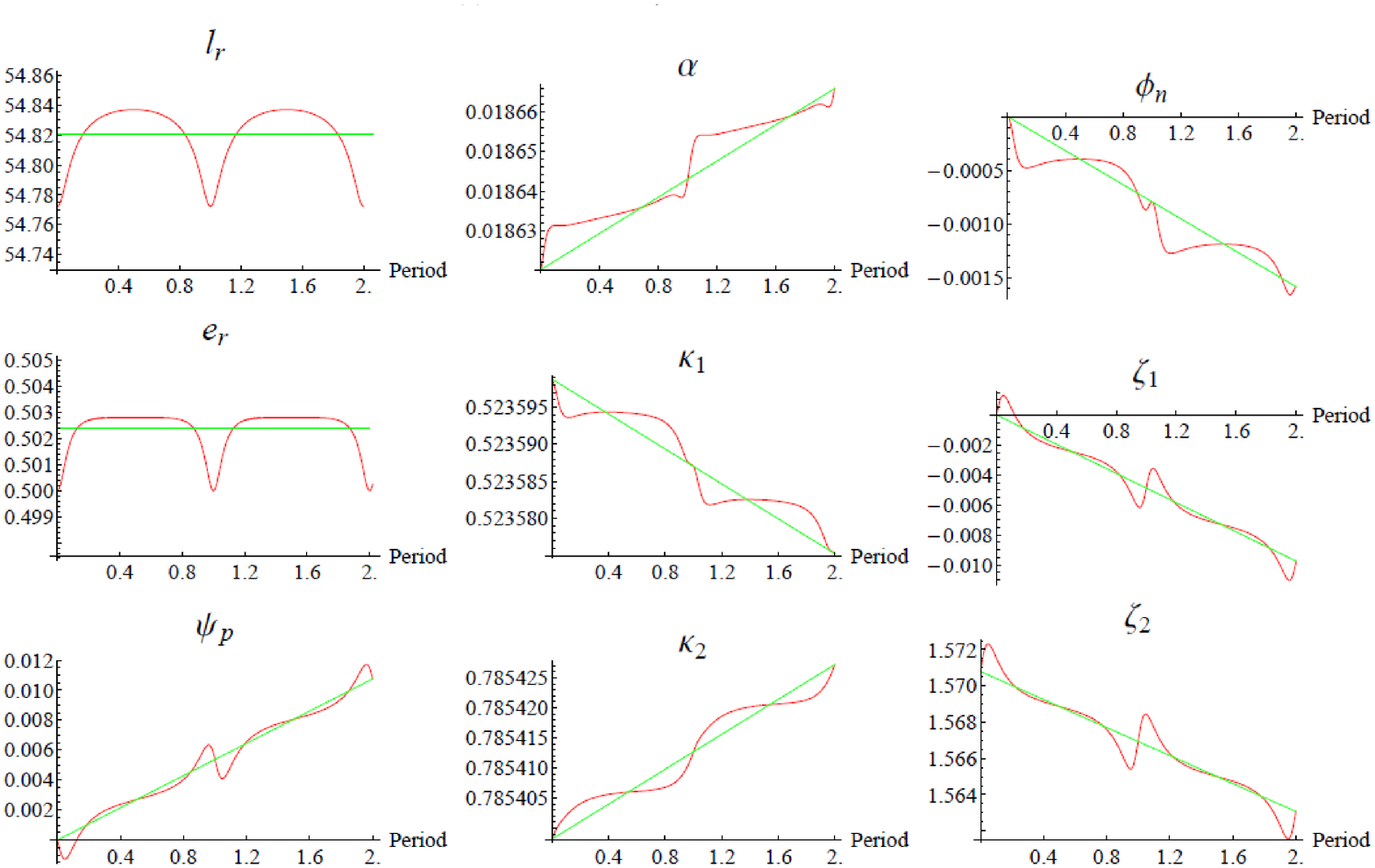}
\end{array}
$

    } \vfill
\subfloat[\large{\textcolor{green}{\line(1,0){15}} Secular, \textcolor{red}{\line(1,0){15}} Instantaneous\label{subfig-4b}}]{$
\begin{array}{ccc}
\includegraphics[scale=0.6]{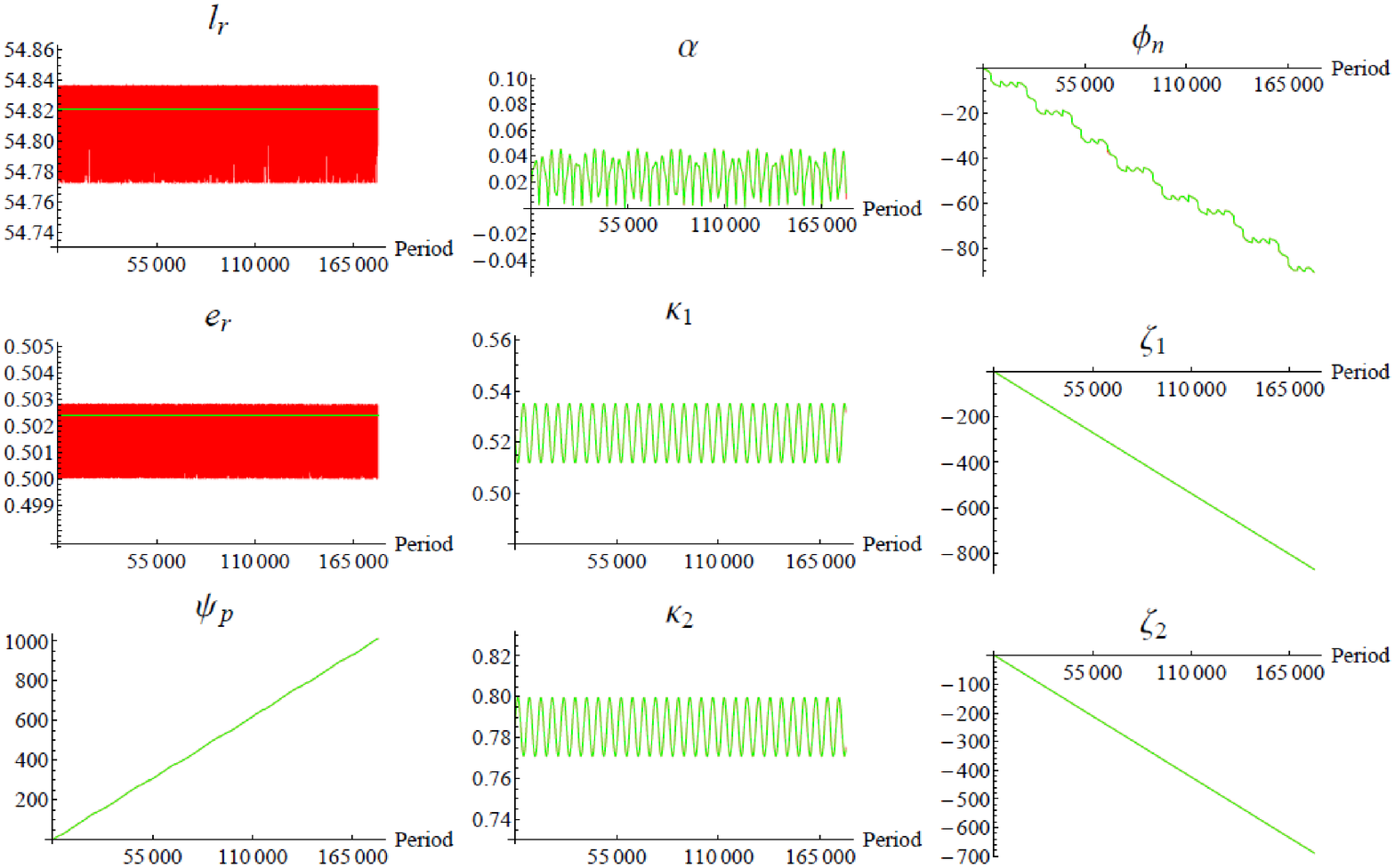}
\end{array}
$
    }
\caption{Comparison of the secular and instantaneous dynamics. The upper
block a) shows the first two periods, while the lower block b) is over the
conservative timescale (shown as the corresponding number of periods on the
x-axis, the conservative timescale being set as $N=0.001\protect\varepsilon %
^{-5/2}$ number of periods). The figures are for total mass $m=20M_{\odot }$%
, mass ratio $\protect\nu =0.5$, dimensionless spin parameters $\protect\chi %
_{1}=0.9982$, $\protect\chi _{2}=0.9982$, PN parameter $\protect\varepsilon %
=0.0005$. The initial values of the other parameters are $e_{r}=0.5$, $%
\protect\kappa _{1}=\protect\pi /6$, $\protect\kappa _{2}=\protect\pi /4$, $%
\protect\zeta _{1}=0$, $\protect\zeta _{2}=\protect\pi /2$, $\protect\psi %
_{p}=0$, $\protect\phi _{n}=0$.}
\label{compfig4}
\end{figure*}

\section{Secular precessing compact binary dynamics\label{secdyn}}

In this section we give the main results of the paper, the averaged
evolution equations of the dimensionless osculating orbital elements $%
\mathfrak{l}_{r}$, $e_{r}$, $\psi _{p}$, $\alpha $, $\phi _{n}$ and spin
angles $\kappa _{i}$, $\zeta _{i}$ ($i=1,2$). These equations contain
post-Newtonian ($PN$), spin-orbit ($SO$), spin-spin ($SS$),
quadrupole-monopole ($QM$) and second post-Newtonian ($2PN$) contributions.

\subsection{Shape variables}

The secular evolution equation of the dimensionless orbital angular momentum 
$\mathfrak{l}_{r}$ reads as 
\begin{equation}
\mathfrak{\bar{\dot{l}}}_{r}=\mathfrak{\bar{\dot{l}}}_{r}^{PN}=\mathfrak{%
\bar{\dot{l}}}_{r}^{SO}=\mathfrak{\bar{\dot{l}}}_{r}^{SS}=\mathfrak{\bar{%
\dot{l}}}_{r}^{QM}=\mathfrak{\bar{\dot{l}}}_{r}^{2PN}=0~.
\end{equation}%
The secular evolution equation of the dimensionless orbital eccentricity $%
e_{r}$ also reads as%
\begin{equation}
\bar{\dot{e}}_{r}=\bar{\dot{e}}_{r}^{PN}=\bar{\dot{e}}_{r}^{SO}=\bar{\dot{e}}%
_{r}^{SS}=\bar{\dot{e}}_{r}^{QM}=\bar{\dot{e}}_{r}^{2PN}=0~.
\end{equation}%
As expected, the average shape of the orbit does not change without
dissipation.

\subsection{Euler angles}

The secular evolution equation of the three Euler angles are given below.

\subsubsection{\protect\bigskip Inclination $\protect\alpha $}

The secular evolution of the inclination $\alpha =\arccos \left( \mathbf{%
\hat{J}}\cdot \mathbf{\hat{L}}_{\mathbf{N}}\right) $ is given by the
following contributions%
\begin{equation}
\bar{\dot{\alpha}}^{PN}=0~,
\end{equation}%
\begin{eqnarray}
\bar{\dot{\alpha}}^{SO} &=&\frac{\eta \pi }{\mathfrak{T}~\mathfrak{\bar{l}}%
_{r}^{3}}\sum_{k=1}^{2}\left( 4\nu ^{2k-3}+3\right)  \notag \\
&&\times \chi _{k}\sin \kappa _{k}\cos \left( \psi _{p}+\zeta _{k}\right) ~,
\label{alphaSO}
\end{eqnarray}%
\begin{eqnarray}
\bar{\dot{\alpha}}^{SS} &=&-\frac{3\eta \pi }{\mathfrak{T~\bar{l}}_{r}^{4}~}%
\chi _{1}\chi _{2}  \notag \\
&&\times \left[ \sin \kappa _{1}\cos \kappa _{2}\cos \left( \psi _{p}+\zeta
_{1}\right) \right.  \notag \\
&&\left. +\cos \kappa _{1}\sin \kappa _{2}\cos \left( \psi _{p}+\zeta
_{2}\right) \right] ~,  \label{alphaSS}
\end{eqnarray}%
\begin{eqnarray}
\bar{\dot{\alpha}}^{QM} &=&-\frac{3\eta \pi }{2\mathfrak{T~\bar{l}}_{r}^{4}~}%
\sum_{k=1}^{2}\nu ^{2k-3}w_{k}\chi _{k}^{2}  \notag \\
&&\times \sin 2\kappa _{k}\cos \left( \psi _{p}+\zeta _{k}\right) ~.
\label{alphaQM}
\end{eqnarray}%
\begin{equation}
\bar{\dot{\alpha}}^{2PN}=0~.
\end{equation}%
As expected, the inclination only changes due to spin and quadrupolar
effects.

\subsubsection{\protect\bigskip The longitude of the ascending node $-%
\protect\phi _{n}$}

The longitude of the ascending node $-\phi _{n}$ is subtended by the
inertial axis $\mathbf{\hat{x}}$ and the ascending node $\mathbf{\hat{l}}=%
\mathbf{\hat{L}}_{\mathbf{N}}\times \mathbf{\hat{J}}$, it has the following
contributions to its secular evolution%
\begin{equation}
\bar{\dot{\phi}}_{n}^{PN}=0~,
\end{equation}%
\begin{eqnarray}
\bar{\dot{\phi}}_{n}^{SO} &=&-\frac{\eta \pi }{\mathfrak{T~\bar{l}}%
_{r}^{3}\sin \alpha }\sum_{k=1}^{2}\left( 4\nu ^{2k-3}+3\right)  \notag \\
&&\times \chi _{k}\sin \kappa _{k}\sin \left( \psi _{p}+\zeta _{k}\right) ~,
\end{eqnarray}%
\begin{eqnarray}
\bar{\dot{\phi}}_{n}^{SS} &=&\frac{3\eta \pi }{\mathfrak{T}~\mathfrak{\bar{l}%
}_{r}^{4}\sin \alpha }\chi _{1}\chi _{2}  \notag \\
&&\times \left[ \sin \kappa _{1}\cos \kappa _{2}\sin \left( \psi _{p}+\zeta
_{1}\right) \right.  \notag \\
&&\left. +\cos \kappa _{1}\sin \kappa _{2}\sin \left( \psi _{p}+\zeta
_{2}\right) \right] ~,
\end{eqnarray}%
\begin{eqnarray}
\bar{\dot{\phi}}_{n}^{QM} &=&\frac{3\eta \pi }{2\mathfrak{T}~\mathfrak{\bar{l%
}}_{r}^{4}\sin \alpha }\sum_{k=1}^{2}\nu ^{2k-3}w_{k}\chi _{k}^{2}  \notag \\
&&\times \sin 2\kappa _{k}\sin \left( \psi _{p}+\zeta _{k}\right) ~,
\end{eqnarray}%
\begin{equation}
\bar{\dot{\phi}}_{n}^{2PN}=0~.
\end{equation}%
Again, only spin and quadrupolar effects contribute.

\subsubsection{\protect\bigskip Argument of the periastron $\protect\psi %
_{p} $}

The secular evolution of the angle $\psi _{p}$ between the node line ($%
\mathbf{\hat{l}}$ perpendicular to both $\mathbf{L}_{\mathbf{N}}$ and $%
\mathbf{J}$) and the periastron ($\mathbf{\hat{A}}_{\mathbf{N}}$) is given by%
\begin{equation}
\bar{\dot{\psi _{p}}}^{PN}=\frac{6\pi }{\mathfrak{T}~\mathfrak{\bar{l}}%
_{r}^{2}}~,  \label{psipPN}
\end{equation}%
\begin{eqnarray}
\bar{\dot{\psi _{p}}}^{SO} &=&-\frac{\eta \pi }{\mathfrak{T}~\mathfrak{\bar{l%
}}_{r}^{3}}\sum_{k=1}^{2}\left( 4\nu ^{2k-3}+3\right)  \notag \\
&&\times \chi _{k}\left[ 2\cos \kappa _{k}\right.  \notag \\
&&\left. +\cot \alpha \sin \kappa _{k}\sin \left( \psi _{p}+\zeta
_{k}\right) \right] ~,  \label{psipSO}
\end{eqnarray}%
\begin{eqnarray}
\bar{\dot{\psi _{p}}}^{SS} &=&\frac{3\eta \pi }{\mathfrak{T}~\mathfrak{\bar{l%
}}_{r}^{4}}\chi _{1}\chi _{2}  \notag \\
&&\times \left\{ \cot \alpha \left[ \sin \kappa _{1}\cos \kappa _{2}\sin
\left( \psi _{p}+\zeta _{1}\right) \right. \right.  \notag \\
&&\left. +\cos \kappa _{1}\sin \kappa _{2}\sin \left( \psi _{p}+\zeta
_{2}\right) \right] +2\cos \kappa _{1}  \notag \\
&&\left. \times \cos \kappa _{2}-\sin \kappa _{1}\sin \kappa _{2}\cos
\,\left( \zeta _{2}-\zeta _{1}\right) \right\} ~,
\end{eqnarray}

\begin{eqnarray}
\bar{\dot{\psi _{p}}}^{QM} &=&\frac{3\eta \pi }{2\mathfrak{T}~\mathfrak{\bar{%
l}}_{r}^{4}}\sum_{k=1}^{2}\nu ^{2k-3}w_{k}\chi _{k}^{2}  \notag \\
&&\times \left[ \cot \alpha \sin 2\kappa _{k}\sin \left( \psi _{p}+\zeta
_{k}\right) \right.  \notag \\
&&\left. -3\sin ^{2}\kappa _{k}+2\right] ~,
\end{eqnarray}%
\begin{equation}
\bar{\dot{\psi _{p}}}^{2PN}=\frac{3\pi }{2\mathfrak{T\bar{l}}_{r}^{4}}\left[
\allowbreak 33\bar{e}_{r}^{2}-4\eta -6\bar{e}_{r}^{2}\eta +2\right] ~.
\end{equation}%
All PN, spin and quadrupolar corrections lead to periastron precession.

\subsection{Spin angles}

The secular evolutions of the spin polar angles $\kappa _{i}$, and azimuthal
angles $\zeta _{i}$ are:%
\begin{equation}
\bar{\dot{\kappa _{i}}}^{PN}=0~,
\end{equation}%
\begin{eqnarray}
\bar{\dot{\kappa _{i}}}^{SO} &=&\frac{\eta \pi }{\mathfrak{T}~\mathfrak{\bar{%
l}}_{r}^{3}}  \notag \\
&&\times \left( 4\nu ^{2j-3}+3\right) \chi _{j}\sin \kappa _{j}\sin \left(
\zeta _{i}-\zeta _{j}\right) ~,
\end{eqnarray}%
\begin{eqnarray}
\bar{\dot{\kappa _{i}}}^{SS} &=&-\frac{\eta \pi }{\mathfrak{T}~\mathfrak{%
\bar{l}}_{r}^{4}}\chi _{j}\sin \kappa _{j}  \notag \\
&&\times \sin \left( \zeta _{i}-\zeta _{j}\right) \left( 2\mathfrak{\bar{l}}%
_{r}\nu ^{2j-3}+3\chi _{i}\cos \kappa _{i}\right) ~,  \label{kappaiSS}
\end{eqnarray}%
\begin{eqnarray}
\bar{\dot{\kappa _{i}}}^{QM} &=&-\frac{3\eta \pi }{2\mathfrak{T}~\mathfrak{%
\bar{l}}_{r}^{4}}\nu ^{2j-3}w_{j}\chi _{j}^{2}  \notag \\
&&\times \sin 2\kappa _{j}\sin \left( \zeta _{i}-\zeta _{j}\right) ~,
\end{eqnarray}%
\begin{equation}
\bar{\dot{\kappa _{i}}}^{2PN}=0~,
\end{equation}%
\begin{equation}
\bar{\dot{\zeta _{i}}}^{PN}=-\bar{\dot{\psi _{p}}}^{PN}~,  \label{zetaiPN}
\end{equation}%
\begin{eqnarray}
\bar{\dot{\zeta _{i}}}^{SO} &=&\frac{\eta \pi }{\mathfrak{T}~\mathfrak{\bar{l%
}}_{r}^{3}}\left\{ \mathfrak{\bar{l}}_{r}\left( 4+3\nu ^{3-2i}\right) \right.
\notag \\
&&+3\left( 4\nu ^{2i-3}+3\right) \chi _{i}\cos \kappa _{i}+\left( 4\nu
^{2j-3}+3\right) \chi _{j}  \notag \\
&&\times \left. \left[ 2\cos \kappa _{j}+\cot \kappa _{i}\sin \kappa
_{j}\cos \left( \zeta _{i}-\zeta _{j}\right) \right] \right\} ~,
\label{zetaiSO}
\end{eqnarray}%
\begin{eqnarray}
\bar{\dot{\zeta _{i}}}^{SS} &=&-\frac{2\eta \pi }{\mathfrak{T}~\mathfrak{%
\bar{l}}_{r}^{3}}\nu ^{2j-3}\chi _{j}\left[ \cos \kappa _{j}\right.  \notag
\\
&&\left. +\cot \kappa _{i}\sin \kappa _{j}\cos \left( \zeta _{i}-\zeta
_{j}\right) \right]  \notag \\
&&-\frac{3\eta \pi }{\mathfrak{T}~\mathfrak{\bar{l}}_{r}^{4}}\chi _{i}\chi
_{j}  \notag \\
&&\times \left\{ \cot \kappa _{i}\left[ 3\sin \kappa _{i}\cos \kappa
_{j}\right. \right.  \notag \\
&&\left. +\cos \kappa _{i}\sin \kappa _{j}\cos \left( \zeta _{i}-\zeta
_{j}\right) \right]  \notag \\
&&\left. -\sin \kappa _{i}\sin \kappa _{j}\cos \left( \zeta _{i}-\zeta
_{j}\right) \right\} ~,  \label{zetaiSS}
\end{eqnarray}%
\begin{eqnarray}
\bar{\dot{\zeta _{i}}}^{QM} &=&-\frac{6\eta \pi }{\mathfrak{T}~\mathfrak{%
\bar{l}}_{r}^{3}}w_{i}\chi _{i}\cos \kappa _{i}  \notag \\
&&-\frac{3\eta \pi }{2\mathfrak{T}~\mathfrak{\bar{l}}_{r}^{4}}%
\sum_{k=1}^{2}w_{k}\nu ^{2k-3}\chi _{k}^{2}  \notag \\
&&\times \left[ \left( 2-3\sin ^{2}\kappa _{k}\right) \right.  \notag \\
&&\left. +\cot \kappa _{i}\sin \left( 2\kappa _{k}\right) \cos \left( \zeta
_{i}-\zeta _{k}\right) \right] ~,
\end{eqnarray}%
\begin{equation}
\bar{\dot{\zeta _{i}}}^{2PN}=-\bar{\dot{\psi _{p}}}^{2PN}~.
\end{equation}%
Here $i\neq j$ and $i=1$, $2$.

One of the advantages of exploring the secular dynamics is that the
evolution of the spin angles decoupled from the rest of the evolutions. The
evolutions for $\kappa _{i}$, $\zeta _{i}$ form a closed set of differential
equations, hence it can be monitored independently.

The average of the precession angular velocities can also be computed from
Eqs (31-33) of Ref. \cite{chameleon}. The expressions are given in Appendix %
\ref{appendix2}.

\begin{figure*}[tbp]
\subfloat[\label{subfig-5a}]{$
\begin{array}{ccc}
\includegraphics[scale=0.66]{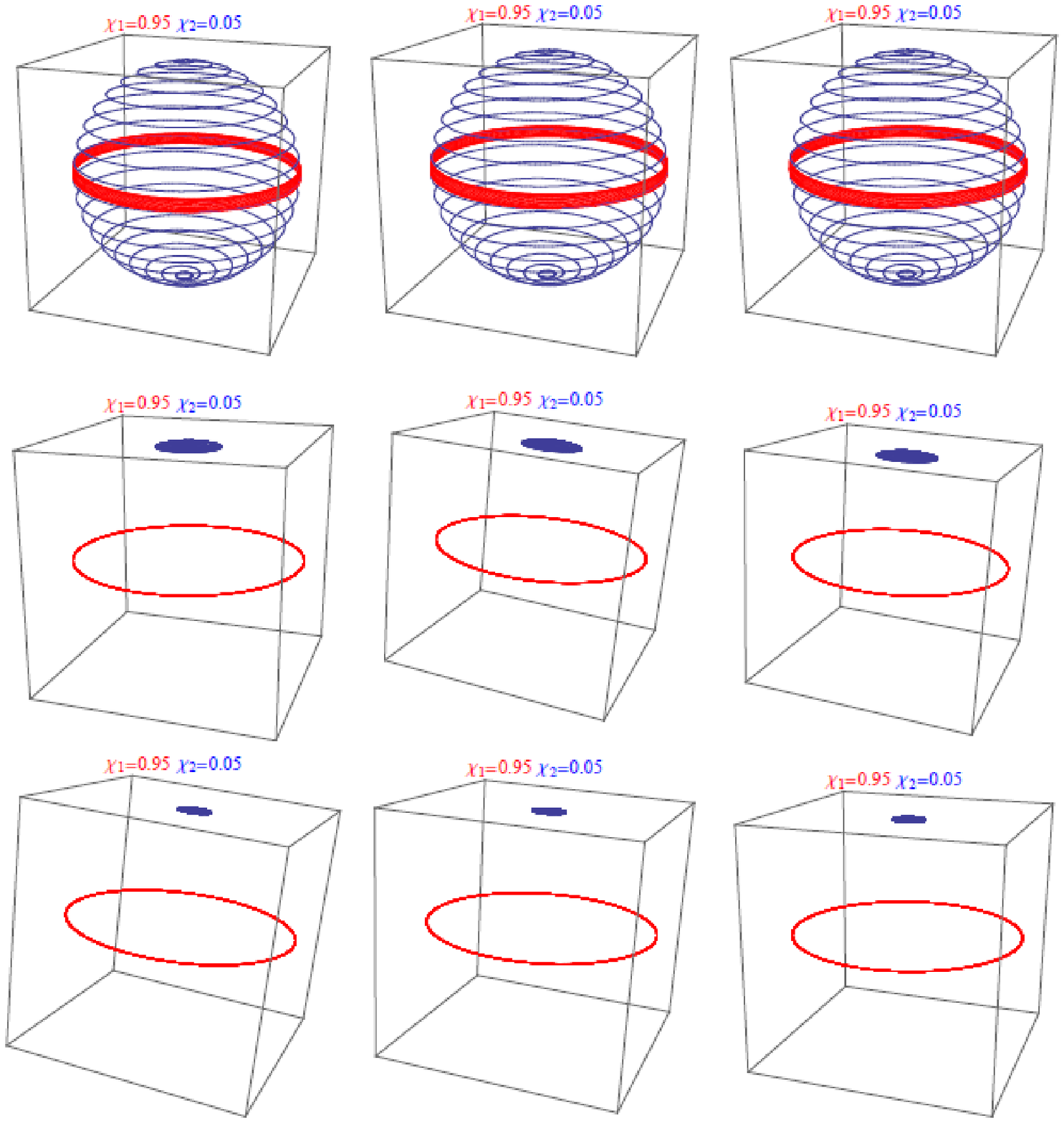} 
\end{array}
$

    } \vfill
\subfloat[\label{subfig-5b}]{$
\begin{array}{ccc}
\includegraphics[scale=0.66]{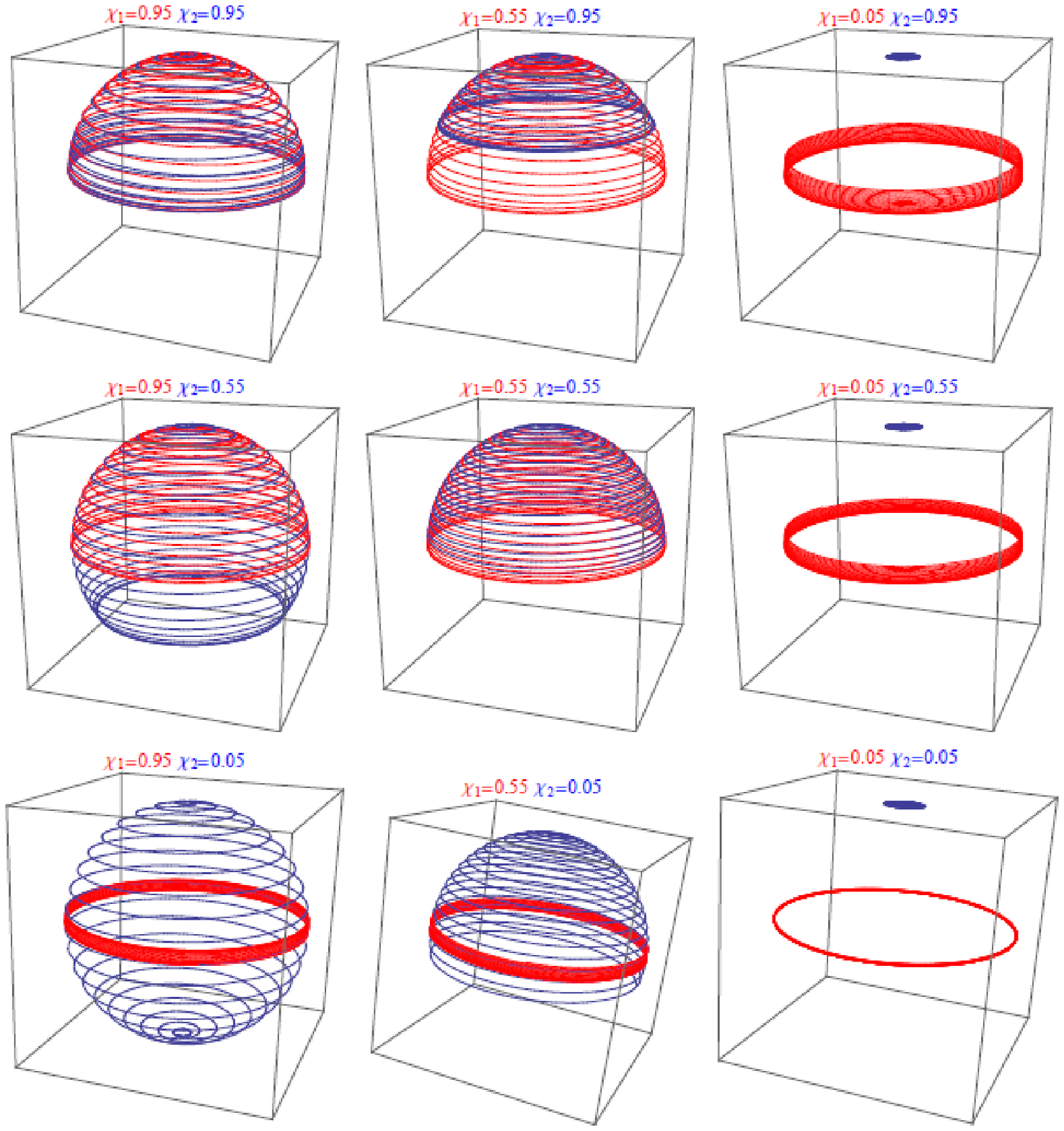}
\end{array}
$
    }
\caption{The upper block a) shows the total mass and mass ratio dependence
of the flip-flop. The total mass changes horizontally as $m=100M_{\odot
},50M_{\odot },10M_{\odot }$, mass ratio changes vertically as $\protect\nu %
=1.0,0.5,0.1$. The dimensionless spin parameters are $\protect\chi _{1}=0.95$%
, $\protect\chi _{2}=0.05$. The lower block b) shows the spin magnitude
dependence of the flip-flop. Horizontally: $\protect\chi _{1}=0.95,0.55,0.05$%
, vertically: $\protect\chi _{2}=0.95,0.55,0.05$. Total mass $m=100M_{\odot
} $ and mass ratio $\protect\nu =1.0$ stay fixed. The PN parameter is set as 
$\protect\varepsilon =0.01$. The other initial values of parameters are: $%
e_{r}=0.1$, $\protect\kappa _{1}=\protect\pi /2-0.001$, $\protect\kappa %
_{2}=0.001$, $\protect\zeta _{1}=0$, $\protect\zeta _{2}=0$, $\protect\psi %
_{p}=0$, $\protect\phi _{n}=0$.}
\label{flipflopfig5}
\end{figure*}

\begin{figure*}[tbp]
\subfloat[\label{subfig-6a}]{$
\begin{array}{ccc}
\includegraphics[scale=0.66]{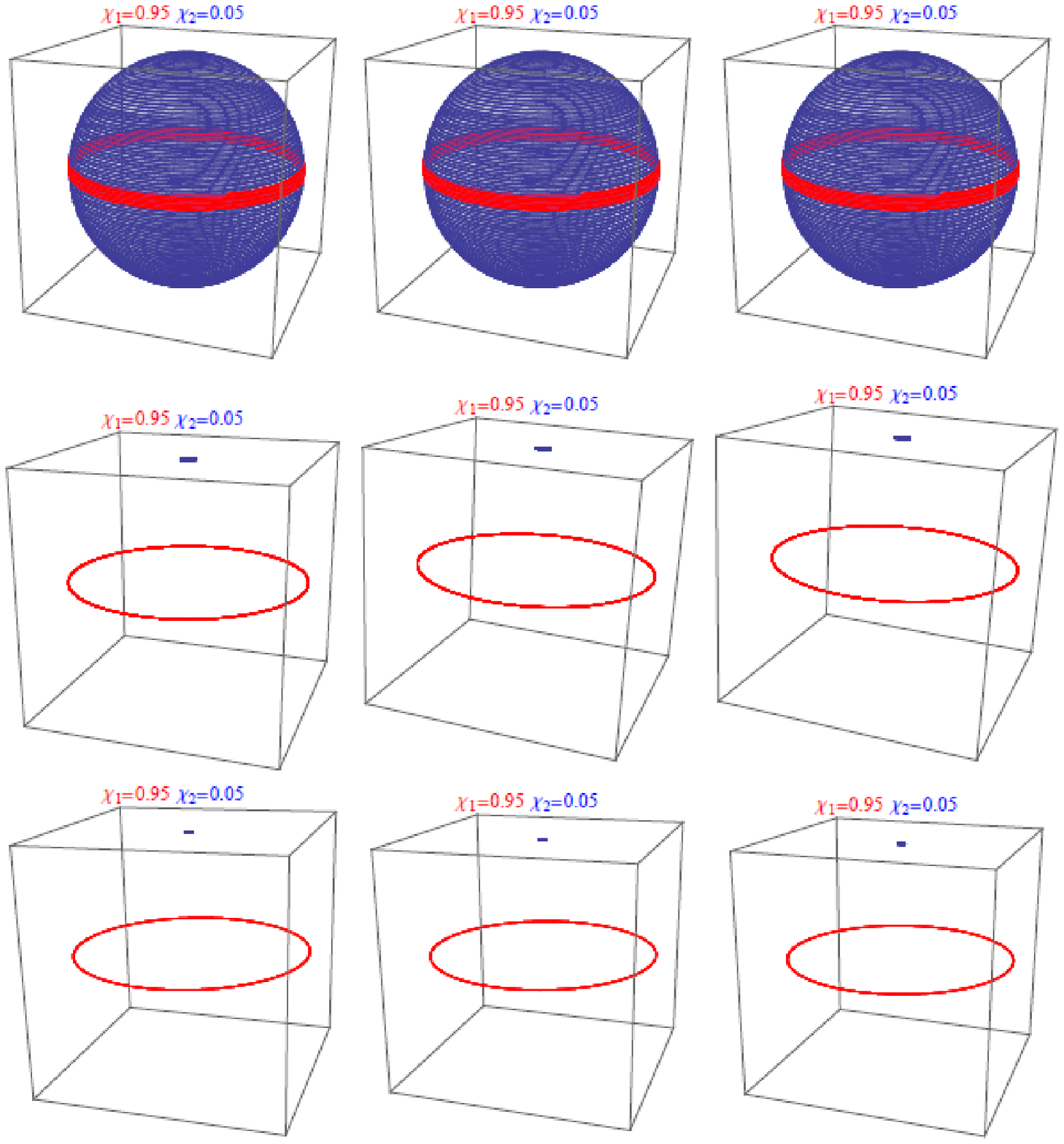}
\end{array}
$

    } \vfill
\subfloat[\label{subfig-6b}]{$
\begin{array}{ccc}
\includegraphics[scale=0.66]{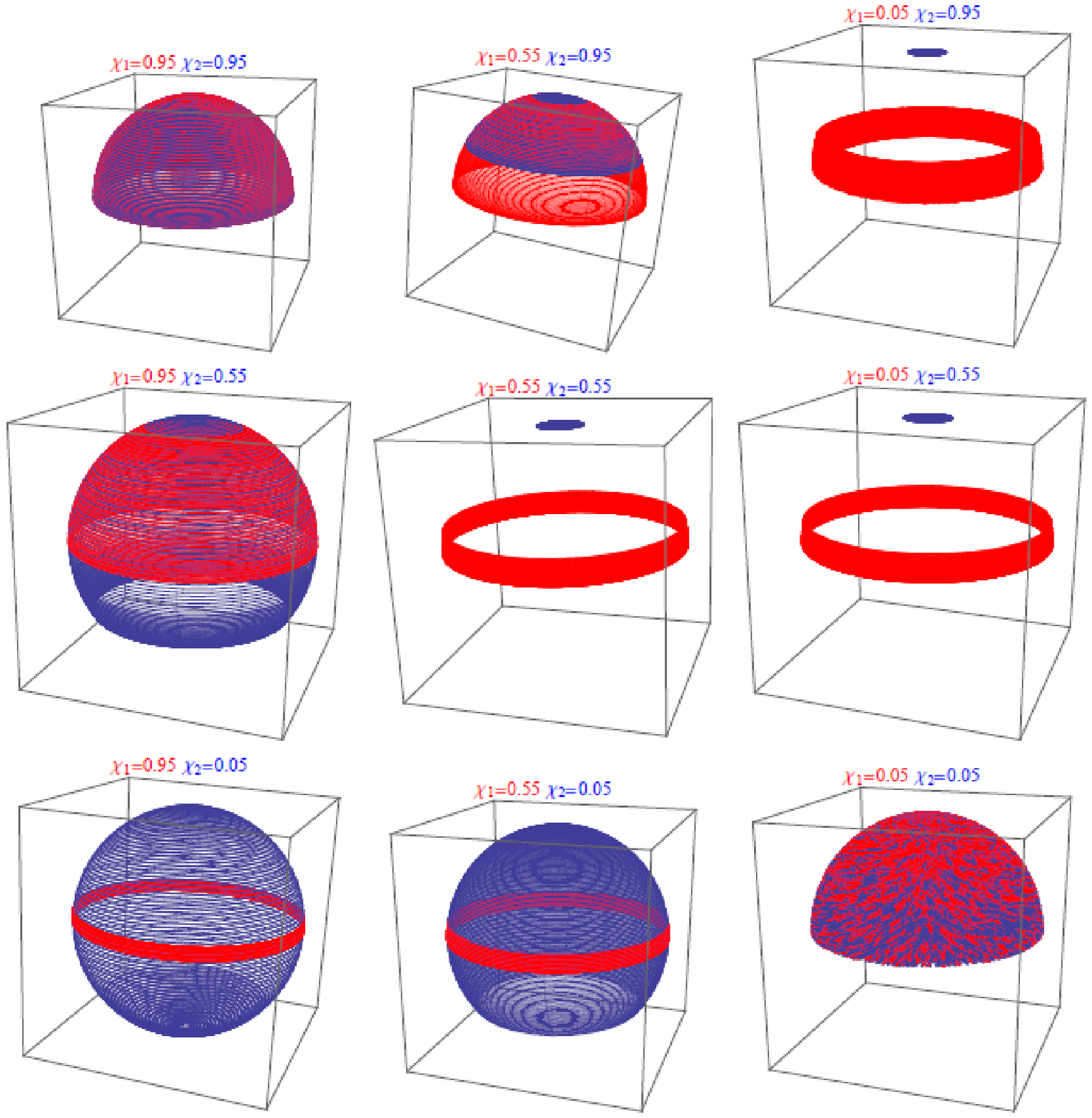}
\end{array}
$
    }
\caption{The upper block a) shows the total mass and mass ratio dependence
of the flip-flop. The total mass changes horizontally as $m=100M_{\odot
},50M_{\odot },10M_{\odot }$, mass ratio changes vertically as $\protect\nu %
=1.0,0.5,0.1$. The dimensionless spin parameters are $\protect\chi _{1}=0.95$%
, $\protect\chi _{2}=0.05$. The lower block b) shows the spin magnitude
dependence of the flip-flop. Horizontally: $\protect\chi _{1}=0.95,0.55,0.05$%
, vertically: $\protect\chi _{2}=0.95,0.55,0.05$. Total mass $m=100M_{\odot
} $ and mass ratio $\protect\nu =1.0$ stay fixed. The PN parameter is set as 
$\protect\varepsilon =0.0005$. The other initial values of parameters are: $%
e_{r}=0.1$, $\protect\kappa _{1}=\protect\pi /2-0.001$, $\protect\kappa %
_{2}=0.001$, $\protect\zeta _{1}=0$, $\protect\zeta _{2}=0$, $\protect\psi %
_{p}=0$, $\protect\phi _{n}=0$.}
\label{flipflopfig6}
\end{figure*}

\begin{figure*}[tbp]
\subfloat[\label{subfig-7a}]{$
\begin{array}{ccc}
\includegraphics[scale=0.66]{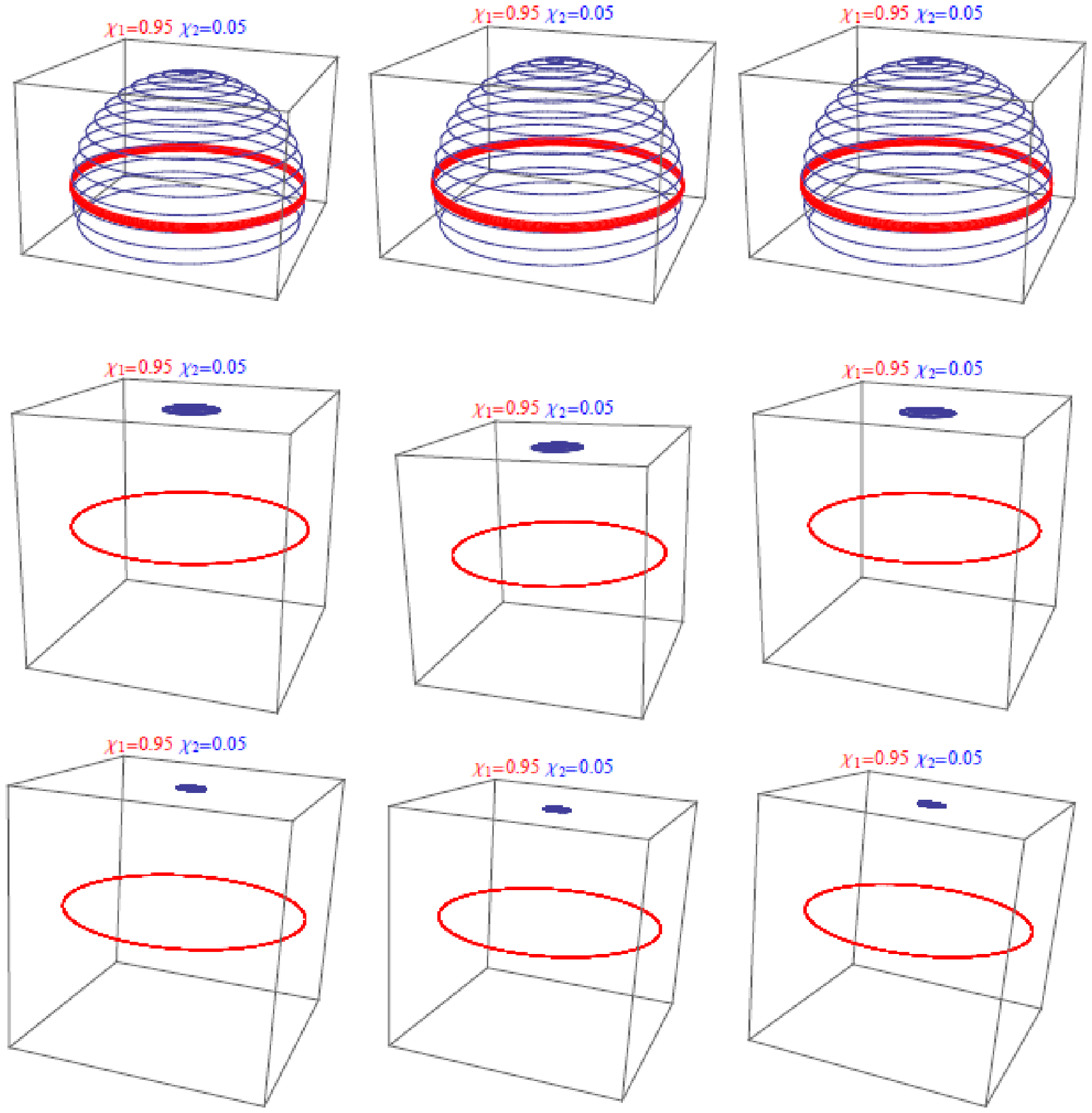} 
\end{array}
$

    } \vfill
\subfloat[\label{subfig-7b}]{$
\begin{array}{ccc}
\includegraphics[scale=0.66]{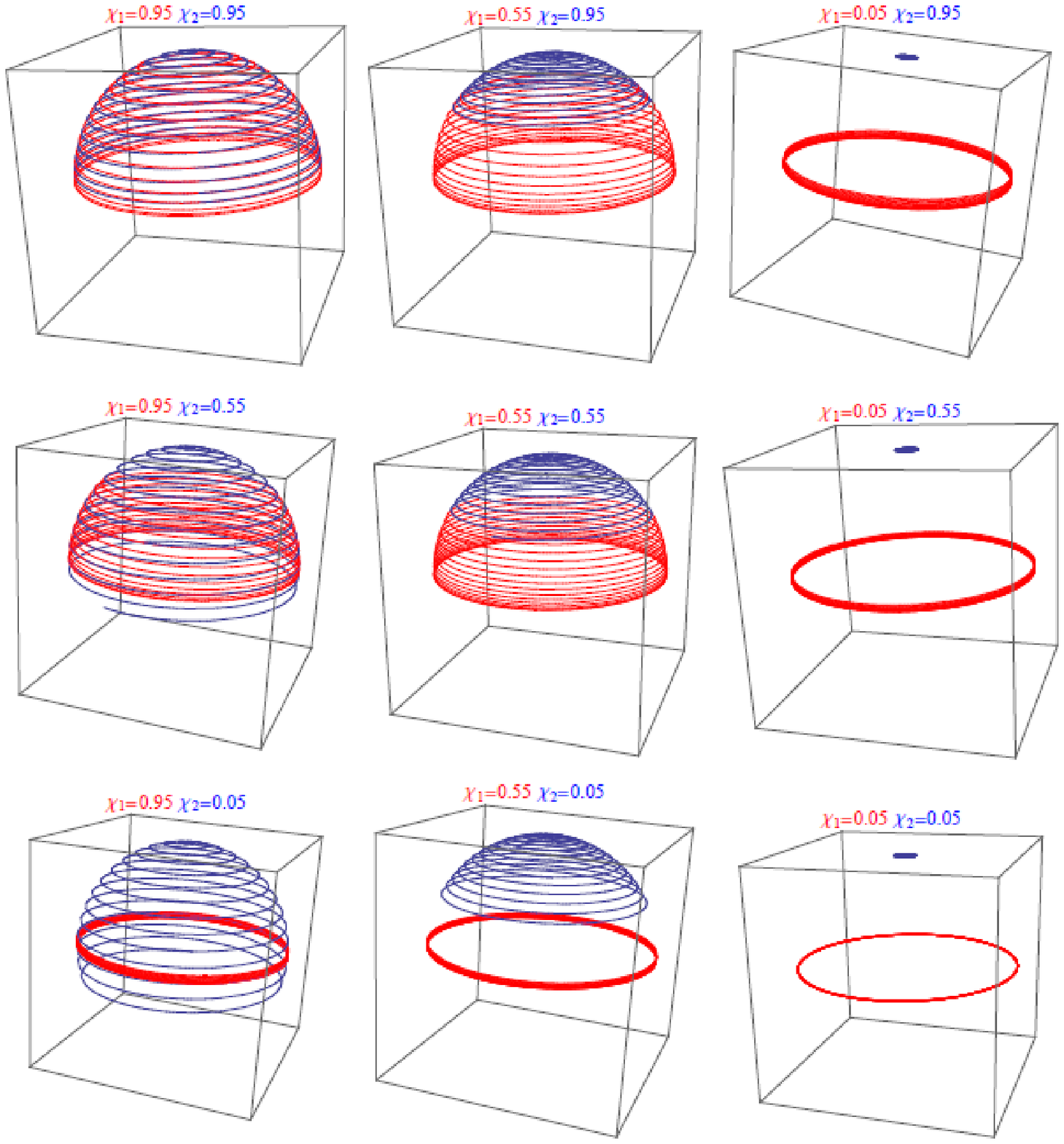}
\end{array}
$
    }
\caption{The upper block a) shows the total mass and mass ratio dependence
of the flip-flop. The total mass changes horizontally as $m=100M_{\odot
},50M_{\odot },10M_{\odot }$, mass ratio changes vertically as $\protect\nu %
=1.0,0.5,0.1$. The dimensionless spin parameters are $\protect\chi _{1}=0.95$%
, $\protect\chi _{2}=0.05$. The lower block b) shows the spin magnitude
dependence of the flip-flop. Horizontally: $\protect\chi _{1}=0.95,0.55,0.05$%
, vertically: $\protect\chi _{2}=0.95,0.55,0.05$. Total mass $m=100M_{\odot
} $ and mass ratio $\protect\nu =1.0$ stay fixed. The PN parameter is set as 
$\protect\varepsilon =0.01$. The other initial values of parameters are: $%
e_{r}=0.5$, $\protect\kappa _{1}=\protect\pi /2-0.001$, $\protect\kappa %
_{2}=0.001$, $\protect\zeta _{1}=0$, $\protect\zeta _{2}=0$, $\protect\psi %
_{p}=0$, $\protect\phi _{n}=0$.}
\label{flipflopfig7}
\end{figure*}

\begin{figure*}[tbp]
\subfloat[\label{subfig-8a}]{$
\begin{array}{ccc}
\includegraphics[scale=0.66]{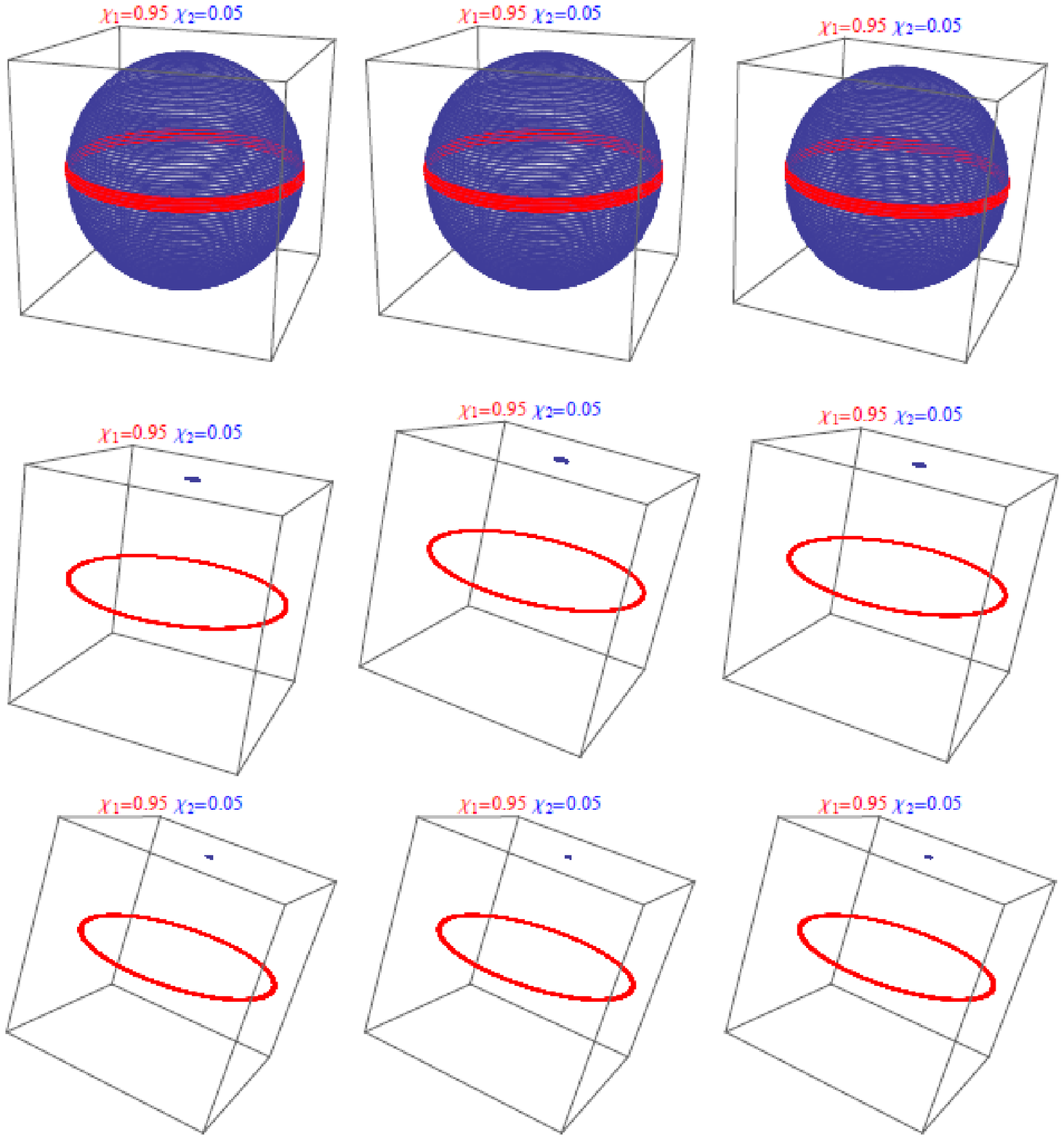} 
\end{array}
$

    } \vfill
\subfloat[\label{subfig-8b}]{$
\begin{array}{ccc}
\includegraphics[scale=0.66]{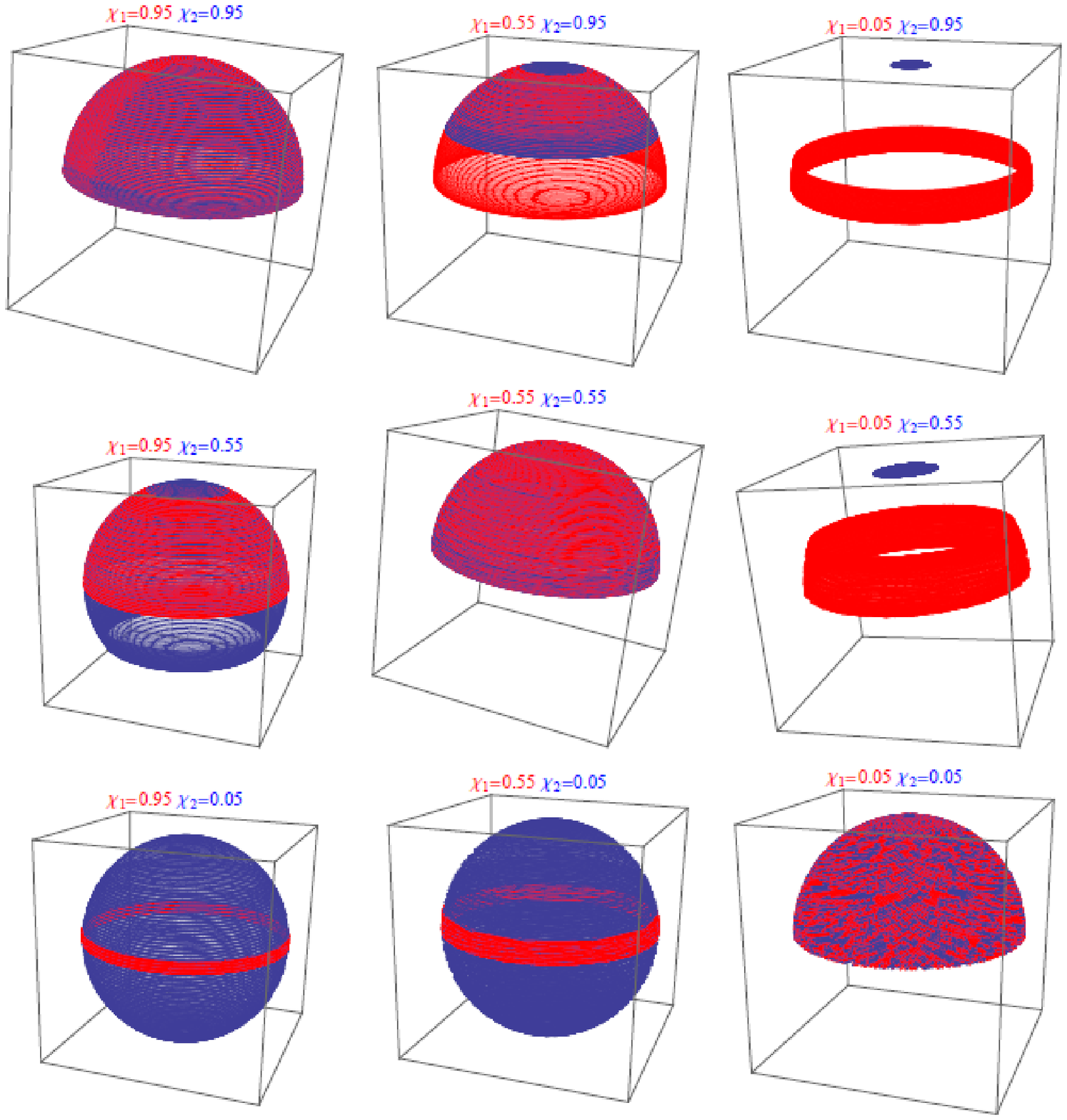} 
\end{array}
$
    }
\caption{The upper block a) shows the total mass and mass ratio dependence
of the flip-flop. The total mass changes horizontally as $m=100M_{\odot
},50M_{\odot },10M_{\odot }$, mass ratio changes vertically as $\protect\nu %
=1.0,0.5,0.1$. The dimensionless spin parameters are $\protect\chi _{1}=0.95$%
, $\protect\chi _{2}=0.05$. The lower block b) shows the spin magnitude
dependence of the flip-flop. Horizontally: $\protect\chi _{1}=0.95,0.55,0.05$%
, vertically: $\protect\chi _{2}=0.95,0.55,0.05$. Total mass $m=100M_{\odot
} $ and mass ratio $\protect\nu =1.0$ stay fixed. The PN parameter is set as 
$\protect\varepsilon =0.0005$. The other initial values of parameters are: $%
e_{r}=0.5$, $\protect\kappa _{1}=\protect\pi /2-0.001$, $\protect\kappa %
_{2}=0.001$, $\protect\zeta _{1}=0$, $\protect\zeta _{2}=0$, $\protect\psi %
_{p}=0$, $\protect\phi _{n}=0$.}
\label{flipflopfig8}
\end{figure*}

\begin{table}[tbp]
\caption{The coefficients $L_{k}$ and $K_{k}$ of $\mathfrak{l}_{r}\left( 
\protect\chi _{p}\right) $.}
\label{tablelrchip}
\begin{center}
\begin{tabular}{cc}
\hline\hline
$Coefficient$ & $Expression$ \\ \hline\hline
$L_{00}^{2PN}$ & $\frac{1}{96}\left[ 432e_{r0}^{3}\eta +e_{r0}^{2}\left(
-117\eta ^{2}+54\eta +48\right) \right. $ \\ 
& $+32e_{r0}\left( 2\eta ^{2}-83\eta +50\right)$ \\ 
& $\left. -48\left( \eta ^{2}-5\eta +6\right) \right]$ \\ \hline
$L_{10}^{2PN}$ & $\frac{e_{r0}}{8}\left[ \left( 2\eta -33\right) \eta
e_{r0}^{2}\right. $ \\ 
& $\left. -116\eta ^{2}+256\eta -160\right]$ \\ \hline
$L_{20}^{2PN}$ & $\frac{1}{8}\left[ e_{r0}^{2}\left( 9\eta ^{2}-3\eta
-4\right) \right.$ \\ 
& $\left. +4\left( \eta ^{2}-5\eta +6\right) \right]$ \\ \hline
$L_{30}^{2PN}$ & $\frac{e_{r0}}{24}\left[ -3\left( 2\eta +3\right) \eta
e_{r0}^{2}\right.$ \\ 
& $\left. +32\eta ^{2}-104\eta +80\right]$ \\ \hline
$L_{40}^{2PN}$ & $\frac{3e_{r0}^{2}(\eta -2)\eta }{32}$ \\ \hline
$L_{01}^{2PN}$ & $\frac{1}{8}\left[ e_{r0}^{2}\left( -9\eta ^{2}+3\eta
+4\right) \right. $ \\ 
& $\left. -4\left( \eta ^{2}-5\eta +6\right) \right] $ \\ \hline
$L_{11}^{2PN}$ & $\frac{e_{r0}}{8}\left[ 3\left( 2\eta +3\right) \eta
e_{r0}^{2}\right. $ \\ 
& $\left. -32\eta ^{2}+104\eta -80\right] $ \\ \hline
$L_{21}^{2PN}$ & $-\frac{9e_{r0}^{2}(\eta -2)\eta }{16} $ \\ \hline
$L_{31}^{2PN}$ & $0 $ \\ \hline
$L_{41}^{2PN}$ & $0 $ \\ \hline
$L_{02}^{2PN}$ & $\frac{3e_{r0}^{2}(\eta -2)\eta }{32} $ \\ \hline
$L_{12}^{2PN}$ & $0 $ \\ \hline
$L_{22}^{2PN}$ & $0$ \\ \hline
$L_{32}^{2PN}$ & $0$ \\ \hline
$L_{42}^{2PN}$ & $0$ \\ \hline\hline
$L_{0}^{SS}$ & $2e_{r0}+3 $ \\ \hline
$L_{1}^{SS}$ & $0 $ \\ \hline
$L_{2}^{SS}$ & $-3 $ \\ \hline
$L_{3}^{SS}$ & $-2e_{r0} $ \\ \hline
$K_{0}^{SS}$ & $-e_{r0} $ \\ \hline
$K_{1}^{SS}$ & $-3 $ \\ \hline
$K_{2}^{SS}$ & $-2e_{r0}$ \\ \hline\hline
$L_{0}^{QM}$ & $-\left( 2e_{r0}+3\right) $ \\ \hline
$L_{1}^{QM}$ & $0 $ \\ \hline
$L_{2}^{QM}$ & $-3 $ \\ \hline
$L_{3}^{QM}$ & $-2e_{r0} $ \\ \hline
$K_{0}^{QM}$ & $-e_{r0} $ \\ \hline
$K_{1}^{QM}$ & $-3 $ \\ \hline
$K_{2}^{QM}$ & $-2e_{r0} $ \\ \hline\hline
\end{tabular}%
\end{center}
\end{table}
\begin{table}[tbp]
\caption{The PN, SS and QM coefficients $E_{k}$ and $F_{k}$ of $e_{r}\left( 
\protect\chi _{p}\right) $.}
\label{tableerchippnssqm}
\begin{center}
\begin{tabular}{cc}
\hline\hline
$Coefficient$ & $Expression$ \\ \hline\hline
$E_{0}^{PN}$ & $3-\eta +\left( 5-4\eta \right) e_{r0}+e_{r0}^{2}\left(
7-6\eta \right)$ \\ \hline
$E_{1}^{PN}$ & $-\left[ 3-\eta +e_{r0}^{2}\left( 7-\frac{11}{2}\eta \right) %
\right] $ \\ \hline
$E_{2}^{PN}$ & $-\left( 5-4\eta \right) e_{r0}$ \\ \hline
$E_{3}^{PN}$ & $\frac{\eta }{2}e_{r0}^{2}$ \\ \hline\hline
$E_{0}^{SS}$ & $-\cos \kappa _{1}\cos \kappa _{2}\left(
e_{r0}^{2}+3e_{r0}+3\right)$ \\ 
& $+\frac{1}{2}\sin \kappa _{1}\sin \kappa _{2}\left[ \left(
e_{r0}^{2}+3e_{r0}+3\right) \cos \zeta _{-}\right.$ \\ 
& $+\left. \left( 7e_{r0}^{2}+15e_{r0}+5\right) \cos \zeta _{+}\right]$ \\ 
\hline
$E_{1}^{SS}$ & $\frac{3}{2}\sin \kappa _{1}\sin \kappa _{2}\left( 3\cos
\zeta _{+}-\cos \zeta _{-}\right) $ \\ 
& $+3\cos \kappa _{1}\cos \kappa _{2}$ \\ \hline
$E_{2}^{SS}$ & $\frac{3}{2}e_{r0}\sin \kappa _{1}\sin \kappa _{2}\left( \cos
\zeta _{+}-\cos \zeta _{-}\right) $ \\ 
& $+3e_{r0}\cos \kappa _{1}\cos \kappa _{2}$ \\ \hline
$E_{3}^{SS}$ & $-\frac{1}{2}e_{r0}^{2}\sin \kappa _{1}\sin \kappa _{2}\left(
\cos \zeta _{+}+\cos \zeta _{-}\right) $ \\ 
& $-7\sin \kappa _{1}\sin \kappa _{2}\cos \zeta _{+} $ \\ 
& $+e_{r0}^{2}\cos \kappa _{1}\cos \kappa _{2} $ \\ \hline
$E_{4}^{SS}$ & $-9e_{r0}\sin \kappa _{1}\sin \kappa _{2}\cos \zeta _{+} $ \\ 
\hline
$E_{5}^{SS}$ & $-3e_{r0}^{2}\sin \kappa _{1}\sin \kappa _{2}\cos \zeta _{+}$
\\ \hline
$F_{0}^{SS}$ & $1-e_{r0}^{2}$ \\ \hline
$F_{1}^{SS}$ & $-3e_{r0}$ \\ \hline
$F_{2}^{SS}$ & $-\left( 2e_{r0}^{2}+7\right)$ \\ \hline
$F_{3}^{SS}$ & $-9e_{r0}$ \\ \hline
$F_{4}^{SS}$ & $-3e_{r0}^{2}$ \\ \hline\hline
$E_{0}^{QM}$ & $\left( 7e_{r0}^{2}+15e_{r0}+5\right) \cos ^{2}\zeta _{i}\sin
^{2}\kappa _{i}$ \\ 
& $+\left( 2e_{r0}^{2}+3e_{r0}-2\right) \cos ^{2}\kappa _{i}$ \\ 
& $-3e_{r0}^{2}-6e_{r0}-1 $ \\ \hline
$E_{1}^{QM}$ & $9\cos ^{2}\kappa _{i}\sin ^{2}\zeta _{i}+3\left( 3\cos
^{2}\zeta _{i}-2\right)$ \\ \hline
$E_{2}^{QM}$ & $3e_{r0}\left( 2-\cos ^{2}\zeta _{i}\right) \cos ^{2}\kappa
_{i}$ \\ 
& $-3e_{r0}\sin ^{2}\zeta _{i}$ \\ \hline
$E_{3}^{QM}$ & $e_{r0}^{2}\cos ^{2}\kappa _{i}$ \\ 
& $+\left[ -\left( e_{r0}^{2}+14\right) \cos ^{2}\zeta _{i}+7\right] \sin
^{2}\kappa _{i}$ \\ \hline
$E_{4}^{QM}$ & $-9e_{r0}\sin ^{2}\kappa _{i}\cos 2\zeta _{i} $ \\ \hline
$E_{5}^{QM}$ & $-3e_{r0}^{2}\sin ^{2}\kappa _{i}\cos 2\zeta _{i}$ \\ \hline
$F_{0}^{QM}$ & $\left( 1-e_{r0}^{2}\right) $ \\ \hline
$F_{1}^{QM}$ & $-3e_{r0}$ \\ \hline
$F_{2}^{QM}$ & $-\left( 2e_{r0}^{2}+7\right)$ \\ \hline
$F_{3}^{QM}$ & $-9e_{r0} $ \\ \hline
$F_{4}^{QM}$ & $-3e_{r0}^{2}$ \\ \hline\hline
\end{tabular}%
\end{center}
\end{table}
\begin{table}[tbp]
\caption{The coefficients $E_{k}^{2PN}$ and $F_{k}^{2PN}$ of $e_{r}\left( 
\protect\chi _{p}\right) $.}
\label{tableerchip2pn}
\begin{center}
\begin{tabular}{cc}
\hline\hline
$Coefficient$ & $Expression$ \\ \hline\hline
$E_{00}^{2PN}$ & $\frac{1}{1920e_{r0}}\left[ 1920e_{r0}^{5}\eta (3\eta
+8)\right. $ \\ 
& $+e_{r0}^{4}\left( -1845\eta ^{2}+8880\eta +1800\right) $ \\ 
& $+32e_{r0}^{3}\left( 232\eta ^{2}-2825\eta +1740\right)$ \\ 
& $-180e_{r0}^{2}\left( 29\eta ^{2}+89\eta +60\right)$ \\ 
& $+160e_{r0}\left( 8\eta ^{2}-187\eta -60\right)$ \\ 
& $\left. -480(\eta -3)^{2}\right] $ \\ \hline
$E_{10}^{2PN}$ & $-\frac{1}{64}\left[ e_{r0}^{4}\eta \left( 161\eta
+477\right) \right. $ \\ 
& $+4e_{r0}^{2}\left( 136\eta ^{2}-849\eta +564\right) $ \\ 
& $\left. +16\eta \left( 8\eta -85\right) \right] $ \\ \hline
$E_{20}^{2PN}$ & $\frac{1}{256e_{r0}}\left[ e_{r0}^{4}\left( 269\eta
^{2}-1312\eta -256\right) \right. $ \\ 
& $+32e_{r0}^{2}\left( 5\eta ^{2}+109\eta +20\right)$ \\ 
& $\left. +64\left( \eta -3\right) ^{2}\right] $ \\ \hline
$E_{30}^{2PN} $ & $\frac{1}{384}\left[ -3e_{r0}^{4}\eta \left( 53\eta
+73\right) \right. $ \\ 
& $+8e_{r0}^{2}\left( 208\eta ^{2}-269\eta +300\right)$ \\ 
& $\left. +128\left( 4\eta ^{2}-17\eta +15\right) \right]$ \\ \hline
$E_{40}^{2PN}$ & $\frac{e_{r0}}{128}\left[ e_{r0}^{2}\left( -13\eta
^{2}+64\eta +8\right) \right.$ \\ 
& $\left. +268\eta ^{2}-676\eta +400\right]$ \\ \hline
$E_{50}^{2PN}$ & $-\frac{3e_{r0}^{2}\eta }{640}\left[ 5e_{r0}^{2}(3\eta
-1)-64\eta +80\right] $ \\ \hline
$E_{60}^{2PN}$ & $\frac{3e_{r0}^{3}\eta ^{2}}{256}$ \\ \hline
$E_{01}^{2PN}$ & $-E_{20}^{2PN}$ \\ \hline
$E_{11}^{2PN}$ & $-3E_{30}^{2PN}$ \\ \hline
$E_{21}^{2PN}$ & $-\frac{3}{2}E_{40}^{2PN}$ \\ \hline
$E_{31}^{2PN}$ & $-\frac{1}{10}E_{50}^{2PN}$ \\ \hline
$E_{41}^{2PN}$ & $-9E_{60}^{2PN}$ \\ \hline
$E_{51}^{2PN}$ & $0$ \\ \hline
$E_{61}^{2PN}$ & $0$ \\ \hline
$E_{02}^{2PN}$ & $E_{40}^{2PN}$ \\ \hline
$E_{12}^{2PN}$ & $\frac{1}{5}E_{50}^{2PN} $ \\ \hline
$E_{22}^{2PN}$ & $9E_{60}^{2PN}$ \\ \hline
$E_{32}^{2PN}$ & $0$ \\ \hline
$E_{42}^{2PN}$ & $0$ \\ \hline
$E_{52}^{2PN}$ & $0$ \\ \hline
$E_{62}^{2PN}$ & $0$ \\ \hline
$E_{03}^{2PN}$ & $-E_{60}^{2PN}$ \\ \hline
$E_{13}^{2PN}$ & $0$ \\ \hline
$E_{23}^{2PN}$ & $0$ \\ \hline
$E_{33}^{2PN}$ & $0$ \\ \hline
$E_{43}^{2PN}$ & $0$ \\ \hline
$E_{53}^{2PN}$ & $0$ \\ \hline
$E_{63}^{2PN}$ & $0$ \\ \hline\hline
\end{tabular}%
\end{center}
\end{table}
\begin{table}[tbp]
\caption{The coefficients $U_{k}$ and $V_{k}$ of $\protect\tau _{02PN}$.}
\label{tabletau02pn}
\begin{center}
\begin{tabular}{cc}
\hline\hline
$Coefficient$ & $Expression$ \\ \hline\hline
$U_{10}$ & $-105\left( \eta +1\right) \eta$ \\ \hline
$U_{9}$ & $10\left( -559\eta +297\eta ^{2}+228\right) $ \\ \hline
$U_{8}$ & $5\left( -2674\eta +1289\eta ^{2}+1336\right) $ \\ \hline
$U_{7}$ & $-4\left( -235\eta +186\eta ^{2}-280\right)$ \\ \hline
$U_{6}$ & $-482\eta ^{2}+25\eta +1120$ \\ \hline
$U_{5}$ & $2\left( -17\,245\eta +5496\eta ^{2}+12\,200\right)$ \\ \hline
$U_{4}$ & $2\left( -2240\eta +79\eta ^{2}+3350\right)$ \\ \hline
$U_{3}$ & $-4\left( -7075\eta +2137\eta ^{2}+5300\right)$ \\ \hline
$U_{2}$ & $2\left( -3915\eta +1317\eta ^{2}+2050\right) $ \\ \hline
$U_{1}$ & $40\left( -254\eta +67\eta ^{2}+210\right)$ \\ \hline
$U_{0}$ & $-20\left( -238\eta +65\eta ^{2}+180\right)$ \\ \hline\hline
$V_{7}$ & $5\left( -932\eta +509\eta ^{2}+504\right)$ \\ \hline
$V_{6}$ & $-5\left( -2700\eta +1399\eta ^{2}+1352\right) $ \\ \hline
$V_{5}$ & $1427\eta ^{2}+920\eta -4240 $ \\ \hline
$V_{4}$ & $12079\eta ^{2}-35880\eta +25680 $ \\ \hline
$V_{3}$ & $-4\left( -6875\eta +2606\eta ^{2}+3650\right) $ \\ \hline
$V_{2}$ & $-4\left( -4735\eta +998\eta ^{2}+4850\right)$ \\ \hline
$V_{1}$ & $40\left( -746\eta +199\eta ^{2}+600\right) $ \\ \hline
$V_{0}$ & $-40\left( -238\eta +65\eta ^{2}+180\right) $ \\ \hline\hline
\end{tabular}%
\end{center}
\end{table}
\begin{table}[tbp]
\caption{The coefficients $\bar{L}_{k}$ and $\bar{K}_{k}$ of $\mathfrak{\bar{%
l}}_{r}$.}
\label{tablelraver}
\begin{center}
\begin{tabular}{cc}
\hline\hline
$Coefficient$ & $Expression$ \\ \hline\hline
$\bar{L}_{4}^{2PN} $ & $15\left( 467\eta ^{2}-580\eta +296\right)$ \\ \hline
$\bar{L}_{3}^{2PN}$ & $480\left( 4\eta ^{2}-3\eta +5\right)$ \\ \hline
$\bar{L}_{2}^{2PN}$ & $-4\left( 3001\eta ^{2}-9445\eta +6610\right)$ \\ 
\hline
$\bar{L}_{1}^{2PN}$ & $-480\left( \eta ^{2}-8\eta +15\right) $ \\ \hline
$\bar{L}_{0}^{2PN}$ & $120\left( 65\eta ^{2}-238\eta +180\right)$ \\ \hline
$\bar{K}_{8}^{2PN}$ & $15\eta (29\eta -3)$ \\ \hline
$\bar{K}_{7}^{2PN}$ & $-60\left( 129\eta ^{2}-188\eta +74\right)$ \\ \hline
$\bar{K}_{6}^{2PN}$ & $-15\left( 116\eta ^{2}-711\eta +304\right)$ \\ \hline
$\bar{K}_{5}^{2PN}$ & $2\left( 5516\eta ^{2}-15155\eta +5420\right) $ \\ 
\hline
$\bar{K}_{4}^{2PN}$ & $-4\left( 5347\eta ^{2}-13720\eta +6400\right) $ \\ 
\hline
$\bar{K}_{3}^{2PN}$ & $8\left( \eta ^{2}+2735\eta -4265\right)$ \\ \hline
$\bar{K}_{2}^{2PN}$ & $20308\eta ^{2}-81610\eta +71380 $ \\ \hline
$\bar{K}_{1}^{2PN}$ & $-720\left( 22\eta ^{2}-82\eta +65\right)$ \\ \hline
$\bar{K}_{0}^{2PN}$ & $60\left( 65\eta ^{2}-238\eta +180\right)$ \\ 
\hline\hline
$\bar{L}^{SS}$ & $4e_{r0}\left( 4e_{r0}^{4}+29e_{r0}^{3}\right.$ \\ 
& $\left. +30e_{r0}^{2}+48e_{r0}+24\right)$ \\ \hline
$\bar{K}^{SS}$ & $\cos \left( \zeta _{1}+\zeta _{2}\right) \left[
96e_{r0}^{3}-236e_{r0}^{2}\right.$ \\ 
& $-171e_{r0}^{4}+95e_{r0}^{5}+56e_{r0}-32 $ \\ 
& $\left. -32\sqrt{1-e_{r0}^{2}}\left( e_{r0}^{2}-e_{r0}^{3}+e_{r0}-1\right) %
\right]$ \\ 
& $-2e_{r0}\cos \left( \zeta _{1}-\zeta _{2}\right) \left(
4e_{r0}^{4}+29e_{r0}^{3}\right. $ \\ 
& $\left. +30e_{r0}^{2}+48e_{r0}+24\right)$ \\ \hline\hline
\end{tabular}%
\end{center}
\end{table}
\begin{table}[tbp]
\caption{The coefficients $\bar{E}_{k}$ and $\bar{F}_{k}$ of $\bar{e}_{r}$.}
\label{tableeraver}
\begin{center}
\begin{tabular}{cc}
\hline\hline
$Coefficient$ & $Expression$ \\ \hline\hline
$\bar{E}_{4}^{2PN} $ & $15\left( 1111\eta ^{2}-1624\eta +528\right)$ \\ 
\hline
$\bar{E}_{3}^{2PN}$ & $4800\left( 4\eta ^{2}-7\eta +4\right)$ \\ \hline
$\bar{E}_{2}^{2PN}$ & $-4\left( 151\eta ^{2}-7550\eta +5640\right) $ \\ 
\hline
$\bar{E}_{1}^{2PN}$ & $2880(\eta -3)\eta $ \\ \hline
$\bar{E}_{0}^{2PN}$ & $8\left( 1501\eta ^{2}-8090\eta +7260\right)$ \\ \hline
$\bar{F}_{8}^{2PN}$ & $120(\eta -3)\eta$ \\ \hline
$\bar{F}_{7}^{2PN}$ & $-60\left( 44\eta ^{2}-107\eta +44\right) $ \\ \hline
$\bar{F}_{6}^{2PN}$ & $-15\left( 317\eta ^{2}-537\eta +96\right)$ \\ \hline
$\bar{F}_{5}^{2PN}$ & $4\left( 457\eta ^{2}-1430\eta -960\right)$ \\ \hline
$\bar{F}_{4}^{2PN}$ & $-5362\eta ^{2}+18335\eta -4380 $ \\ \hline
$\bar{F}_{3}^{2PN}$ & $-6\left( 714\eta ^{2}-4315\eta +5420\right)$ \\ \hline
$\bar{F}_{2}^{2PN}$ & $21\left( 391\eta ^{2}-2110\eta +2100\right)$ \\ \hline
$\bar{F}_{1}^{2PN}$ & $-4\left( 1411\eta ^{2}-7820\eta +7260\right)$ \\ 
\hline
$\bar{F}_{0}^{2PN}$ & $1501\eta ^{2}-8090\eta +7260$ \\ \hline\hline
$\bar{E}^{SS}$ & $16e_{r0}^{3}(1+e_{r0})^{2}\sqrt{1-e_{r0}^{2}}\left(
e_{r0}^{2}+2\right)$ \\ \hline
$\bar{F}^{SS}$ & $-\left[ 8(e_{r0}+1)^{2}\sqrt{1-e_{r0}^{2}}\right. $ \\ 
& $\times \left( e_{r0}^{2}+2\right) e_{r0}^{3}\cos (\zeta _{1}-\zeta _{2})$
\\ 
& $+\cos \left( \zeta _{1}+\zeta _{2}\right) \left[
3176e_{r0}^{3}-3176e_{r0}^{2}\right.$ \\ 
& $+1552e_{r0}^{4}-1552e_{r0}^{5}-35e_{r0}^{6}$ \\ 
& $+\sqrt{1-e_{r0}^{2}}\left( 2376e_{r0}^{2}-2328e_{r0}^{3}\right.$ \\ 
& $-468e_{r0}^{4}+606e_{r0}^{5}-35e_{r0}^{6}$ \\ 
& $\left. \left. +92e_{r0}^{7}+1600e_{r0}-1600\right) \right]$ \\ 
\hline\hline
\end{tabular}%
\end{center}
\end{table}
\begin{table}[tbp]
\caption{The coefficients $L_{0}$ and $K_{0}$.}
\label{lr0kifcoeffs}
\begin{center}
\begin{tabular}{cc}
\hline\hline
$Coefficient$ & $Expression$ \\ \hline\hline
$L_{0,2}^{2PN}$ & $3\left( 75\eta ^{2}-176\eta +216\right)$ \\ \hline
$L_{0,1}^{2PN}$ & $48\left( 3\eta ^{2}-11\eta +10\right)$ \\ \hline
$L_{0,0}^{2PN}$ & $86\eta ^{2}-260\eta +176 $ \\ \hline
$K_{0,4}^{2PN}$ & $12\left( 12\eta ^{2}-39\eta +28\right)$ \\ \hline
$K_{0,3}^{2PN}$ & $-6\left( 18\eta ^{2}-63\eta +64\right)$ \\ \hline
$K_{0,2}^{2PN}$ & $3\left( 17\eta ^{2}-7\eta +28\right)$ \\ \hline
$K_{0,1}^{2PN}$ & $-2\left( 7\eta ^{2}+2\eta -32\right)$ \\ \hline
$K_{0,0}^{2PN}$ & $43\eta ^{2}-130\eta +88$ \\ \hline\hline
$L_{0}^{SS}$ & $4\bar{e}_{r}\left( 4\bar{e}_{r}^{4}+53\bar{e}_{r}^{3}\right.$
\\ 
& $\left. +78\bar{e}_{r}^{2}+72\bar{e}_{r}+24\right)$ \\ \hline
$K_{0}^{SS}$ & $\left[ -95\bar{e}_{r}^{5}+171\bar{e}_{r}^{4}-96\bar{e}%
_{r}^{3}\right.$ \\ 
& $+236\bar{e}_{r}^{2}-56\bar{e}_{r}+32$ \\ 
& $\left. -32\left( 1-\bar{e}_{r}\right) \left( 1-\bar{e}_{r}^{2}\right)
^{3/2}\right]$ \\ 
& $\times \cos (\zeta _{1}+\zeta _{2})+96\bar{e}_{r}^{2} $ \\ 
& $\times \left( 1+\bar{e}_{r}\right) ^{2}\left( 2\cos \zeta _{1}\cos \zeta
_{2}\right.$ \\ 
& $\left. -\sin \zeta _{1}\sin \zeta _{2}\right) +2\bar{e}_{r}\left( 4\bar{e}%
_{r}^{4}\right.$ \\ 
& $\left. +29\bar{e}_{r}^{3}+30\bar{e}_{r}^{2}+48\bar{e}_{r}+24\right) $ \\ 
& $\times \cos (\zeta _{1}-\zeta _{2})$ \\ \hline\hline
\end{tabular}%
\end{center}
\end{table}
\begin{table}[tbp]
\caption{The coefficients $L_{0}$ and $K_{0}$.}
\label{er0kifcoeffs}
\begin{center}
\begin{tabular}{cc}
\hline\hline
$Coefficient$ & $Expression$ \\ \hline\hline
$E_{0,4}^{2PN}$ & $15\left( 2915\eta ^{2}-8904\eta +6192\right)$ \\ \hline
$E_{0,3}^{2PN}$ & $960\left( 36\eta ^{2}-102\eta +65\right)$ \\ \hline
$E_{0,2}^{2PN}$ & $4\left( 7673\eta ^{2}-20110\eta +15360\right)$ \\ \hline
$E_{0,1}^{2PN}$ & $960\left( 2\eta ^{2}-11\eta +15\right)$ \\ \hline
$E_{0,0}^{2PN}$ & $-8\left( 4559\eta ^{2}-13390\eta +9540\right)$ \\ \hline
$F_{0,6}^{2PN}$ & $15\left( 383\eta ^{2}-989\eta +560\right)$ \\ \hline
$F_{0,5}^{2PN}$ & $-30\left( 51\eta ^{2}-261\eta +224\right)$ \\ \hline
$F_{0,4}^{2PN}$ & $-15\left( 68\eta ^{2}-301\eta +232\right)$ \\ \hline
$F_{0,3}^{2PN}$ & $2\left( 893\eta ^{2}-3505\eta +3360\right)$ \\ \hline
$F_{0,2}^{2PN}$ & $-2\left( 1756\eta ^{2}-6425\eta +5250\right)$ \\ \hline
$F_{0,1}^{2PN}$ & $9358\eta ^{2}-28100\eta +20880$ \\ \hline
$F_{0,0}^{2PN}$ & $-4559\eta ^{2}+13390\eta -9540$ \\ \hline\hline
$E_{0}^{SS}$ & $16\bar{e}_{r}^{3}\left( 1+\bar{e}_{r}\right) \left( \bar{e}%
_{r}^{3}+3\bar{e}_{r}^{2}+4\bar{e}_{r}+2\right)$ \\ \hline
$F_{0}^{SS}$ & $\cos \left( \zeta _{1}+\zeta _{2}\right) \left[ 92\bar{e}%
_{r}^{7}-35\bar{e}_{r}^{6}\right.$ \\ 
& $+606\bar{e}_{r}^{5}-468\bar{e}_{r}^{4}-2328\bar{e}_{r}^{3}$ \\ 
& $+2376\bar{e}_{r}^{2}+1600\bar{e}_{r}-1600 $ \\ 
& $\left. +8\left( \bar{e}_{r}+1\right) \left( 3\bar{e}_{r}^{2}+200\right)
\left( 1-\bar{e}_{r}\right) ^{5/2}\right]$ \\ 
& $+32(\bar{e}_{r}+1)^{2}\bar{e}_{r}^{4} $ \\ 
& $\times \left( 2\cos \zeta _{1}\cos \zeta _{2}-\sin \zeta _{1}\sin \zeta
_{2}\right)$ \\ 
& $+8(\bar{e}_{r}+1)^{2}\left( \bar{e}_{r}^{2}+2\right) \bar{e}_{r}^{3} $ \\ 
& $\times \cos (\zeta _{1}-\zeta _{2})$ \\ \hline\hline
\end{tabular}%
\end{center}
\end{table}

\section{Limits of validity\label{limits}}

\subsection{Validity of the conservative approach: the conservative timescale%
}

We evolve the conservative secular dynamics up to the point where the
neglected 2.5PN radiation reaction term would generate an error of $p\%$,
and choose the number $p$ according to the desired accuracy. This is
described with the $N$ number of periods to evolve, when the error
accumulating from neglecting the radiation reaction reaches $p\%$:%
\begin{equation}
N=\frac{p}{100}\varepsilon ^{-5/2}~.  \label{constimescale}
\end{equation}%
The number of periods $N$ defines the conservative timescale. In what
follows, we proceed with the choices $p=1$ or $p=0.1$, depending on the
circumstances.

\subsection{Accuracy of the secular dynamics on precessional timescale and
above, up to the conservative timescale}

We check the long-term accuracy of the secular dynamics by a numerical
comparison with instantaneous dynamics, as given by Eqs. (36)-(42) of Ref. 
\cite{chameleon}. The results are represented on Figs. \ref{compfig1}-\ref%
{compfig4}.

On each figure the upper block a) shows the first two periods of the
evolution, and the lower block b) shows the evolution of the variables over
the conservative timescale.

On Figs. \ref{compfig1} and \ref{compfig2} the evolution is presented for $%
p=1$, thus for $N=0.01\varepsilon ^{-5/2} $. On Fig. \ref{compfig1} the PN
parameter and the eccentricity are $\varepsilon =0.01$ and $e_{r}=0.1$
respectively. On Fig. \ref{compfig2} (as compared to Fig. \ref{compfig1})
only the eccentricity is changed to $e_{r}=0.5$. These figures show that the
both spin axes are more stable for larger eccentricity. This can be seen
from the angles $\kappa _{1}$, $\kappa _{2}$, $\zeta _{1}$ and $\zeta _{2}$
on each figure's block a) within one period. Block b) also shows this over
the conservative timescale. The plane of the motion changes less as shown by
the angle $\alpha $. The reason for this is that for lower eccentricity the
compact objects stay close to each other over a larger part of the period,
thus have more time to interact. This is ensured by the initial conditions
being set such that at the periastron in both cases the distance between the
compact objects is the same.

On Figs. \ref{compfig3} and \ref{compfig4} the evolution is presented for $%
p=0.1$ only, thus for $N=0.001\varepsilon ^{-5/2} $. Figs. \ref{compfig3}
and \ref{compfig4} show the previous eccentricity values for the PN
parameter value $\varepsilon =0.0005$. These figures exhibit the same
differences for the two eccentricities. For the decrease in the PN
parameter, the number of precessional periods increases over the
conservative timescale. For low $\varepsilon $ values long term precessional
effects may be easier to identify.

\section{Spin flip-flops\label{spinflipflop}}

In this Section we analyse the flip-flop effect for various parameter
ranges. This effect has been shown \cite{flipflop,flipflop1,flipflop2} by
numerical investigations for particular configurations (equal masses, the
dominant spin in the plane of motion and the smaller spin perpendicular to
it) and it was claimed that to some extent it also occurs for generic mass
ratios.

For our analysis we keep the specific spin orientations, but we vary the
total mass, mass ratio, and spin magnitudes in each of the Figures \ref%
{flipflopfig5}-\ref{flipflopfig8}. The $z$-axis is given by the Newtonian
orbital angular momentum. While this direction is not conserved (the
direction of the total angular momentum $\mathbf{J}$ is) on all figures the
parameters are such that $\mathbf{L}_{\mathbf{N}}$ is the dominant
contribution to $\mathbf{J}$ (the spin magnitudes are smaller than $L_{N}$).
Hence the evolution is represented in a quasi-inertial system.

On all Figs. \ref{flipflopfig5}-\ref{flipflopfig8} the upper block $a)$
shows the total mass and mass ratio dependence and the lower block $b)$
shows the dimensionless spin magnitude dependence of the flip-flop
phenomenon. For all Figs. the evolution is presented for $p=1$. On Figs. \ref%
{flipflopfig5} and \ref{flipflopfig6} the eccentricity is set as $\bar{e}%
_{r}=0.1$ and the PN parameter changes as $\varepsilon =0.01$ and $%
\varepsilon =0.0005$\ respectively. On Figs. \ref{flipflopfig7} and \ref%
{flipflopfig8} the eccentricity is set as $\bar{e}_{r}=0.5$ and the PN
parameter changes as $\varepsilon =0.01$ and $\varepsilon =0.0005$\
respectively.

The upper blocks shows that the flip-flop phenomenon slightly depends on the
eccentricity. For larger eccentricity under the same time period the $\kappa
_{2}$ angle changes less, the number of precessional period it takes for the
full flip-flop to happen is larger. The reason for this is that for small
eccentricity, the compact objects spend more time close to each other over
one period, compared to larger eccentricities. For smaller PN parameter
value however the flip-flop effect happens several times on the conservative
timescale, and several full flip-flop can occur for larger eccentricities.
The upper panels also show that the effect does not depend on the total
mass. The effect actually depends heavily on the mass ratio. For equal
masses, the effect is prominent, however as we start to decrease the mass
ratio, the effect disappears.

The lower blocks of Figs. \ref{flipflopfig5}-\ref{flipflopfig8} the effect
of change in both dimensionless spin parameters $\chi _{1}$ and $\chi _{2}$
is shown. For equal spin magnitude values both spin angles $\kappa _{1}$ and 
$\kappa _{2}$ change back and forth by $\pi /2$. As we decrease the spin
magnitude that closely aligns the orbital angular momentum, the flip-flop
effect appears, and the larger the ratio of $\chi _{1}/\chi _{2}$ is, the
more prominent the flip-flop effect is. If we change the ratio in the
opposite direction the flip-flop effect completely disappears. When the
ratio is $1$, the aligned spin moves into the plane of the motion, and the
spin initially in the plane of the motion becomes aligned with the orbital
angular momentum. Than they keep changing place back and forth over the
conservative timescale. The change of the PN parameter and the eccentricity
has the same effect as for the total mass and mass ratio dependence.

This systematic mapping of the parameter space $\left( m,\nu ,\chi _{1},\chi
_{2},\varepsilon ,\overline{e}_{r}\right) $ indicates that the spin
flip-flop appears only at equal and co-measurable masses. The effect is too
small to manifest on the conservative timescale for other parameter ranges.

\section{ Orbital angular momentum flip-flop\label{lnflipflop}}

\begin{figure}[tbp]
$%
\begin{array}{ccc}
\includegraphics[scale=0.6]{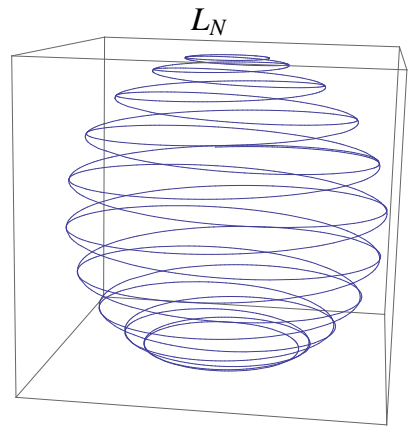} &  & 
\end{array}%
$%
\caption{The flip-flop of the orbital angular momentum with $\protect\alpha $
the polar angle and $\protect\psi _{p}$ the azimuthal angle. Over several
precession periods the plane of the motion flips upside down.}
\label{lnflipflopfig1}
\end{figure}

\begin{figure*}[tbp]
$%
\begin{array}{ccc}
\includegraphics[scale=0.5]{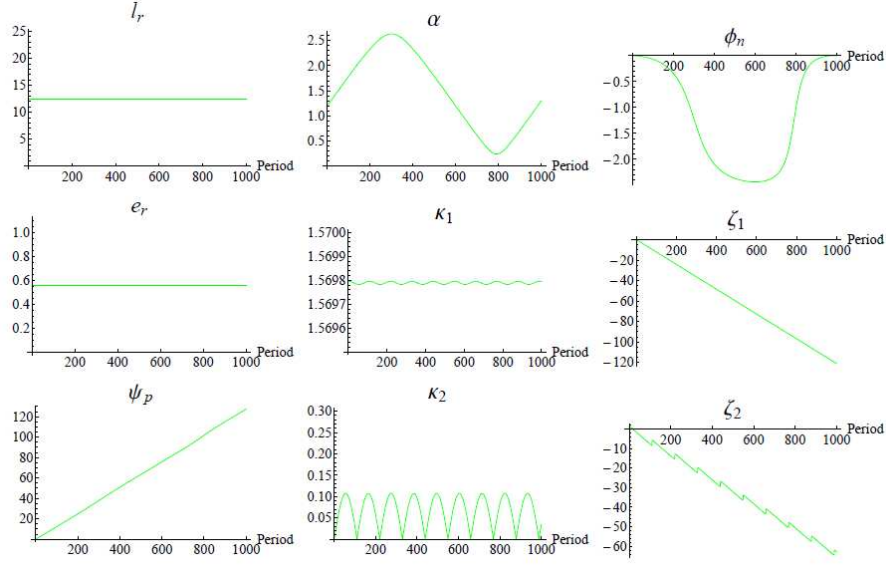} 
\end{array}
$%
\caption{The dynamics of the system where the orbital angular momentum
undergoes a flip-flop like behaviour is shown over the conservative
timescale with $p=1$. The parameters for the system are as follows: total
mass $m = 100 M_{\odot}$, mass ratio $\protect\nu = 0.03$, dimensionless
spin parameters $\protect\chi_{1} = 0.95$, $\protect\chi_{2} = 0.05$ . Value
of the PN parameter $\protect\varepsilon = 0.01$. Initial values of the
parameters: $e_{r}=0.5$, $\protect\kappa_{1}=\protect\pi/2-0.001$, $\protect%
\kappa_{2}=0.001$, $\protect\zeta_{1}=0$, $\protect\zeta_{2}=\protect\pi/2$, 
$\protect\psi_{p}=0$, $\protect\phi_{n}=0$.}
\label{lnflipflopfig2}
\end{figure*}

The derived secular dynamics is suitable to easily monitor the evolution of
the spin and orbital angular momentum vectors over several precessional
cycles up to the conservative timescale. The peculiar evolution of the spins
was discussed in the previous section. In this Section we give an example
when the orbital angular momentum vector undergoes a similar flip-flopping
behaviour. In this example the larger spin dominates over the orbital
angular momentum, $L_{N}/S_{1}\approx 0.3$. We represent the temporal
evolution of $\mathbf{L}_{\mathbf{N}}$ in the system with $\mathbf{J}$ on
the $z$-axis. For convenience the role of time is taken by the periastron
precession angle $\psi _{p}$. The venturing of the orbital angular momentum
vector from pole to pole through several precession cycles in represented on
Fig. \ref{lnflipflopfig1}, where the $\alpha $ and $\psi _{p}$ angles are
taken as the polar and azimuthal angle. On Fig. \ref{lnflipflopfig2} the
secular evolution of the dynamics of this particular system is shown. 

\section{Concluding Remarks\label{concludingr}}

In this paper we derived the conservative secular evolution of precessing
compact binaries, to second post-Newtonian order accuracy, with
leading-order spin-orbit, spin-spin and mass quadrupole-monopole
contributions included. The secular evolution equations emerged as a closed
system of first-order differential equations, which in contrast with the
instantaneous dynamics, is autonomous. The dependent variables are the polar
angles of the spin vectors $\kappa _{1}$ and $\kappa _{2}$, the azimuthal
angle of the spin vectors $\zeta _{1}$ and $\zeta _{2}$, the angles $\alpha $
and $\phi _{n}$ giving the orientation of the orbital angular momentum
vector, together with the periastron angle $\psi _{p}$, the dimensionless
magnitude of the orbital angular momentum $\mathfrak{l}_{r}$ and the
eccentricity $e_{r}$.

We proved that the secular dynamics reliably characterizes the system over
the conservative timescales. The later was defined as in Eq. (\ref%
{constimescale}), typically larger than the precessional timescale, however
shorter than the timescale of gravitational radiation. We compared the
secular and instantaneous dynamics over this timescale.

The analytic equations are suitable to monitor the spin flip-flop effect,
recently found by numerical relativity methods. Our investigations showed
that the effect does not generalize beyond its original parameter settings:
the dominant spin is in the plane of the motion, the smaller flip-flopping
spin is initially closely aligned to the orbital angular momentum and the
system has equal mass ratio.

We investigated the PN parameter and eccentricity dependence of the
flip-flop effect on Figs \ref{flipflopfig5}-\ref{flipflopfig8}. The
parameters were set as $\varepsilon =0.01,0.0005$ and $e_{r}=0.1,0.5$. When
the PN parameter decreases, the flip-flop effect occurs several times over
the conservative timescale. Due to the compact objects staying together for
larger part of the period for small eccentricity than for large, the
frequency of the flip-flop effect is higher for small eccentricities.

As shown on the upper block a) of Figs. \ref{flipflopfig5}-\ref{flipflopfig8}
the flip-flop phenomenon does not depend on the total mass, but
significantly depends on the mass ratio, preferring close to equal mass
systems. The lower block b) of the same Figs showed that the effect depends
on the ratio of the dimensionless spin magnitudes. The effect is most
prominent when the spin almost aligned with the orbital angular momentum is
small, and the spin in the plane of the motion dominates. When the ratio of
the spin magnitudes is $1$, both spins do a half flip-flop over the
conservative timescale.

In addition, a similar evolution of the orbital angular momentum vector was
found, which ventures from one pole to another through several precessional
periods. For this phenomenon to occur the larger spin has to dominate over
the orbital angular momentum. In our investigation we found the effect for
the ratio $L_{N}/S_{1}\approx 0.3$. This can only be achieved with small
mass ratios $\nu \ll 1$ during the inspiral.

\section{Acknowledgements}

In the early stages of this work L. \'{A}. G. was supported by the European
Union and the State of Hungary, co-financed by the European Social Fund in
the framework of T\'{A}MOP 4.2.4. A/2-11-/1-2012-0001 \textquotedblleft
National Excellence Program.\textquotedblright

\appendix

\section{The $\protect\chi_{p}$ dependent shape variables and the period 
\label{appendix1}}

We derive the $\chi _{p}$ dependence of the dimensionless orbital angular
momentum $\mathfrak{l}_{r}\left( \chi _{p}\right) $ and the dimensionless
orbital eccentricity $e_{r}\left( \chi _{p}\right) $. A lower index $0$
indicates values taken at $\chi _{p}=0$.%
\begin{eqnarray}
\mathfrak{l}_{r}\left( \chi _{p}=0\right) &=&\mathfrak{l}_{r0}~, \\
e_{r}\left( \chi _{p}=0\right) &=&e_{r0}~.
\end{eqnarray}%
With $\dot{\chi}_{p}$ containing Newtonian order terms, to calculate the
period to 2PN order, the $\chi _{p}$ dependence of $\mathfrak{l}_{r}\left(
\chi _{p}\right) $ and of $e_{r}\left( \chi _{p}\right) $ has to be taken
into account up to 2PN order. The rest of the orbital elements $\psi _{p}$, $%
\alpha $, $\phi _{n}$, and spin angles $\kappa _{i}$, $\zeta _{i}$ ($i=1,2$)
do not enter at Newtonian order, hence for the 2PN calculations, only their
leading order expressions are needed, which are constant.

\subsection{The $\protect\chi _{p}$ dependence of $\mathfrak{l}_{r}\label%
{lrevo}$}

Formal integration of Eqs. (36) of Ref. \cite{chameleon} by exploring it's
Eq. (43) gives 
\begin{eqnarray}
\mathfrak{l}_{r}\left( \chi _{p}\right) &=&\mathfrak{l}_{r0}+\int_{0}^{\chi
_{p}}\frac{\dot{l}_{r}}{\dot{\chi}_{p}}d\chi _{p}  \notag \\
&=&\mathfrak{l}_{r0}+\frac{1}{\mathfrak{l}_{r0}}\mathfrak{l}_{rPN}\left(
\chi _{p}\right) +\frac{1}{\mathfrak{l}_{r0}^{2}}\mathfrak{l}_{rSO}\left(
\chi _{p}\right)  \notag \\
&&+\frac{1}{\mathfrak{l}_{r0}^{3}}\mathfrak{l}_{rQM}\left( \chi _{p}\right) +%
\frac{1}{\mathfrak{l}_{r0}^{3}}\mathfrak{l}_{rSS}\left( \chi _{p}\right) 
\notag \\
&&+\frac{1}{\mathfrak{l}_{r0}^{3}}\mathfrak{l}_{r2PN}\left( \chi _{p}\right)
~,
\end{eqnarray}%
with

\begin{equation}
\mathfrak{l}_{rPN}\left( \chi _{p}\right) =2\left( 2-\eta \right)
e_{r0}\left( 1-\cos \chi _{p}\right) ~,
\end{equation}%
\begin{equation}
\mathfrak{l}_{r2PN}\left( \chi _{p}\right)
=\sum_{k=0}^{4}\sum_{l=0}^{2}L_{kl}^{2PN}\sin ^{2l}\chi _{p}\cos ^{k}\chi
_{p}~,
\end{equation}%
\begin{eqnarray}
\mathfrak{l}_{rSO}\left( \chi _{p}\right) &=&-\frac{\eta e_{r0}}{2}\left(
\cos \chi _{p}-1\right)  \notag \\
&&\times \sum_{k=1}^{2}\left( 4^{2k-3}+3\right) \chi _{k}\cos \kappa _{k}~,
\end{eqnarray}%
\begin{eqnarray}
\mathfrak{l}_{rSS}\left( \chi _{p}\right) &=&\eta \chi _{1}\chi _{2}\sin
\kappa _{1}\sin \kappa _{2}  \notag \\
&&\times \left[ \cos \zeta _{+}\sum_{k=0}^{3}L_{k}^{SS}\cos ^{k}\chi
_{p}\right.  \notag \\
&&\left. +\sin \zeta _{+}\sin \chi _{p}\sum_{k=0}^{2}K_{k}^{SS}\cos ^{k}\chi
_{p}\right] ~,
\end{eqnarray}%
\begin{eqnarray}
\mathfrak{l}_{rQM}\left( \chi _{p}\right) &=&\frac{\eta }{2}%
\sum_{i=1}^{2}\chi _{i}^{2}w_{i}\nu ^{2i-3}\sin ^{2}\kappa _{i}  \notag \\
&&\times \left[ \cos 2\zeta _{i}\sum_{k=0}^{3}L_{k}^{QM}\cos ^{k}\chi
_{p}\right.  \notag \\
&&\left. +\sin 2\zeta _{i}\sin \chi _{p}\sum_{k=0}^{2}K_{k}^{QM}\cos
^{k}\chi _{p}\right] ~,
\end{eqnarray}%
where the coefficients $L_{k}$ and $K_{k}$ of $\mathfrak{l}_{r}\left( \chi
_{p}\right) $ are shown in Table \ref{tablelrchip}.

\subsection{The $\protect\chi _{p}$ dependence of $e_{r}$ $\label{erevo}$}

Formal integration of Eqs. (37) of Ref. \cite{chameleon} by exploring it's
Eq. (43) leads to%
\begin{eqnarray}
e_{r}\left( \chi _{p}\right) &=&e_{r0}+\int_{0}^{\chi _{p}}\frac{\dot{e}_{r}%
}{\dot{\chi}_{p}}d\chi _{p}~=  \notag \\
&=&e_{r0}+\frac{1}{\mathfrak{l}_{r0}^{2}}e_{rPN}\left( \chi _{p}\right) +%
\frac{1}{\mathfrak{l}_{r0}^{3}}e_{rSO}\left( \chi _{p}\right)  \notag \\
&&+\frac{1}{\mathfrak{l}_{r0}^{4}}e_{rQM}\left( \chi _{p}\right) +\frac{1}{%
\mathfrak{l}_{r0}^{4}}e_{rSS}\left( \chi _{p}\right)  \notag \\
&&+\frac{1}{\mathfrak{l}_{r0}^{4}}e_{r2PN}\left( \chi _{p}\right) ~,
\end{eqnarray}%
with%
\begin{equation}
e_{rPN}\left( \chi _{p}\right) =\sum_{k=0}^{3}E_{k}^{PN}\cos ^{k}\chi _{p}~,
\end{equation}%
\begin{eqnarray}
e_{rSO}\left( \chi _{p}\right) &=&\frac{\eta }{2}\left( 1-e_{r0}^{2}\right)
\left( 1-\cos \chi _{p}\right) \times  \notag \\
&&\times \sum_{i=1}^{2}\left( 4^{2k-3}+3\right) \chi _{i}\cos \kappa _{i}~,
\end{eqnarray}%
\begin{gather}
e_{rSS}\left( \chi _{p}\right) =\eta \chi _{1}\chi _{2}\left[
\sum_{k=0}^{5}E_{k}^{SS}\cos ^{k}\chi _{p}\right.  \notag \\
\left. +\sin \kappa _{1}\sin \kappa _{2}\sin \zeta _{+}\sin \chi
_{p}\sum_{k=0}^{4}F_{k}^{SS}\cos ^{k}\chi _{p}\right] ~,
\end{gather}%
\begin{eqnarray}
e_{rQM}\left( \chi _{p}\right) &=&\frac{\eta }{2}\sum_{i=1}^{2}\chi
_{i}^{2}\nu ^{2i-3}w_{i}  \notag \\
&&\times \left[ \sum_{k=0}^{6}E_{k}^{QM}\cos ^{k}\chi _{p}\right.  \notag \\
&&+\sin 2\zeta _{i}\sin ^{2}\kappa _{i}\sin \chi _{p}  \notag \\
&&\left. \times \sum_{k=0}^{4}F_{k}^{QM}\cos ^{k}\chi _{p}\right] ~,
\end{eqnarray}%
The coefficients $E_{k}$ and $F_{k}~$of $e_{r}\left( \chi _{p}\right) $ for
the PN, SS\ and QM contributions are shown in Table \ref{tableerchippnssqm}.
~%
\begin{equation}
e_{r2PN}\left( \chi _{p}\right)
=\sum_{l=0}^{3}\sum_{k=0}^{6}E_{kl}^{2PN}\cos ^{k}\chi _{p}\sin ^{2l}\chi
_{p}~.
\end{equation}%
The coefficients $E_{k}^{2PN}$ and $F_{k}^{2PN}~$of $e_{r}\left( \chi
_{p}\right) $ are shown in Table \ref{tableerchip2pn}.

\subsection{Dimensionless 2PN radial period}

Everything required to calculate the 2PN radial period is given now. We
insert the expressions of $\mathfrak{l}_{r}\left( \chi _{p}\right) $\ and $%
e_{r}\left( \chi _{p}\right) $ into the integral (\ref{perioddef}) and
Taylor-expand to 2PN order. The various contributions to the period (\ref%
{period}) read as%
\begin{equation}
\mathfrak{T}_{0}=\frac{2\pi \mathfrak{l}_{r0}^{3}}{\left(
1-e_{r0}^{2}\right) ^{3/2}}~,  \label{tau0expr}
\end{equation}%
\begin{equation*}
\tau _{01PN}=-\frac{\left( 1-e_{r0}^{2}\right) \left( e_{r0}^{2}(7\eta
-6)+2e_{r0}(5\eta -3)+4\eta -18\right) }{2(e_{r0}-1)^{2}}~,
\end{equation*}%
\begin{equation}
\tau _{0SO}=0~,
\end{equation}%
\begin{eqnarray}
\tau _{02PN} &=&\frac{1}{40(1-e_{r0})^{2}e_{r0}^{4}}\left[
\sum_{k=0}^{10}U_{k}e_{r0}^{k}\right.  \notag \\
&&\left. -\frac{\left( 1-e_{r0}^{2}\right) ^{3/2}}{2(1-e_{r0})}%
\sum_{k=0}^{7}V_{k}e_{r0}^{k}\right] ~,
\end{eqnarray}%
\begin{eqnarray}
\tau _{0SS} &=&-\frac{3\chi _{1}\chi _{2}\left( 1+e_{r0}\right) ^{2}\eta }{%
\left( 1-e_{r0}\right) \mathfrak{l}_{r0}^{4}}\left[ \cos \kappa _{1}\cos
\kappa _{2}+\sin \kappa _{1}\sin \kappa _{2}\right.  \notag \\
&&\left. \times \left( \sin \zeta _{1}\sin \zeta _{2}-2\cos \zeta _{1}\cos
\zeta _{2}\right) \right] ~,
\end{eqnarray}%
\begin{eqnarray}
\tau _{0QM} &=&-\frac{3\eta \left( 1+e_{r0}\right) ^{2}}{2\left(
1-e_{r0}\right) \mathfrak{l}_{r0}^{4}}\sum_{i=1}^{2}\chi _{i}^{2}\nu
^{2i-3}w_{i}  \notag \\
&&\times \left( 1-3\sin ^{2}\kappa _{i}\cos ^{2}\zeta _{i}\right) ~.
\end{eqnarray}%
The coefficients $U_{k}$ and $V_{k}$ of $\tau _{02PN}$ are enlisted in Table %
\ref{tabletau02pn}.

\section{The secular shape variables $\mathfrak{\bar{l}}$ and $\bar{e}_{r}$ 
\label{lreraver}}

We calculate the secular shape variables. For $\mathfrak{\bar{l}}_{r}$\ we
compute the integral given in Eq. (\ref{lraverdefkif}). We get 
\begin{equation}
\mathfrak{\bar{l}}_{rN}=\frac{2\pi \mathfrak{l}_{r0}^{4}}{\left(
1-e_{r0}^{2}\right) ^{3/2}}~,  \label{lravernewtexpr}
\end{equation}%
\begin{eqnarray}
\mathfrak{\bar{l}}_{rPN} &=&-\frac{\pi \mathfrak{l}_{r0}^{2}}{(1-e_{r0})^{2}%
\sqrt{1-e_{r0}^{2}}}\left[ e_{r0}^{2}(3\eta +2)\right.  \notag \\
&&\left. +14e_{r0}(\eta -1)+4\eta -18\right] ~,
\end{eqnarray}%
\begin{eqnarray}
\mathfrak{\bar{l}}_{rSO} &=&-\frac{\pi e_{r0}\eta \mathfrak{l}_{r0}}{%
(1-e_{r0})\sqrt{1-e_{r0}^{2}}}  \notag \\
&&\times \sum_{k=1}^{2}\left( 4^{2k-3}+3\right) \chi _{k}\cos \kappa _{k}~,
\end{eqnarray}%
\begin{eqnarray}
\mathfrak{\bar{l}}_{r2PN} &=&\frac{\pi }{120e_{r0}^{4}}\sum_{k=0}^{4}\bar{L}%
_{k}^{2PN}e_{r0}^{k}  \notag \\
&&-\frac{\pi \sqrt{1-e_{r0}^{2}}}{60(e_{r0}-1)^{4}e_{r0}^{4}}\sum_{k=0}^{8}%
\bar{K}_{k}^{2PN}e_{r0}^{k}~,
\end{eqnarray}%
\begin{eqnarray}
\mathfrak{\bar{l}}_{rSS} &=&\frac{\pi \chi _{1}\chi _{2}\eta }{16\left(
1-e_{r0}\right) e_{r0}^{2}\left( 1-e_{r0}^{2}\right) ^{3/2}}  \notag \\
&&\times \left( \cos \kappa _{1}\cos \kappa _{2}\bar{L}^{SS}\right.  \notag
\\
&&\left. +\sin \kappa _{1}\sin \kappa _{2}\bar{K}^{SS}\right) ~,
\end{eqnarray}%
\begin{eqnarray}
\mathfrak{\bar{l}}_{rQM} &=&\frac{\pi \left( e_{r0}^{2}+2\right) \eta }{%
256e_{r0}\left( 1-e_{r0}^{2}\right) ^{5/2}}\sum_{k=1}^{2}\chi _{k}^{2}\nu
^{2k-3}w_{k}  \notag \\
&&\times \left[ -4\left( 47e_{r0}^{3}+1050e_{r0}^{2}+488e_{r0}\right. \right.
\notag \\
&&\left. +480\right) \sin ^{2}\kappa _{k}\cos 2\zeta _{k}  \notag \\
&&+16\left( 5e_{r0}^{3}+21e_{r0}^{2}+15e_{r0}+6\right)  \notag \\
&&\left. \times \left( 3\cos 2\kappa _{k}+1\right) \right] ~.
\end{eqnarray}%
Here the coefficients $\bar{L}_{k}$ and $\bar{K}_{k}$ of $\mathfrak{\bar{l}}%
_{r}$ are shown in Table \ref{tablelraver}.

For $\mathfrak{\bar{l}}_{r}$\ we compute the integral given in Eq. (\ref%
{eraverdefkif}). We get%
\begin{equation}
\bar{e}_{rN}=\frac{2\pi e_{r0}\mathfrak{l}_{r0}^{3}}{\left(
1-e_{r0}^{2}\right) ^{3/2}}~,  \label{eravernewtexpr}
\end{equation}%
\begin{eqnarray}
\bar{e}_{rPN} &=&\frac{\pi \mathfrak{l}_{r0}}{e_{r0}}\left\{ 2(3\eta
-5)\right.  \notag \\
&&+\frac{\sqrt{1-e_{r0}^{2}}}{(1-e_{r0})^{3}(e_{r0}+1)}\left[
4e_{r0}^{4}\left( \eta -2\right) \right.  \notag \\
&&-20e_{r0}^{3}\left( \eta -1\right) +e_{r0}^{2}\left( 22-9\eta \right) 
\notag \\
&&\left. \left. +2e_{r0}\left( 5\eta -7\right) -6\eta +10\right] \right\} ~,
\end{eqnarray}%
\begin{eqnarray}
\bar{e}_{rSO} &=&\frac{\pi (e_{r0}+1)\eta }{\sqrt{1-e_{r0}^{2}}}  \notag \\
&&\times \sum_{k=1}^{2}\left( 4^{2k-3}+3\right) \chi _{k}\cos \kappa _{k}~,
\end{eqnarray}%
\begin{eqnarray}
\bar{e}_{r2PN} &=&\frac{\pi }{480e_{r0}^{3}\mathfrak{l}_{r0}}\sum_{k=0}^{4}%
\bar{E}_{k}^{2PN}e_{r0}^{k}  \notag \\
&&-\frac{\pi \sqrt{1-e_{r0}^{2}}}{60(e_{r0}-1)^{4}e_{r0}^{3}\mathfrak{l}_{r0}%
}\sum_{k=0}^{8}\bar{F}_{k}^{2PN}e_{r0}^{k}~,
\end{eqnarray}%
\begin{eqnarray}
\bar{e}_{rSS} &=&\frac{3\pi \chi _{1}\chi _{2}\eta }{16\left(
1-e_{r0}\right) e_{r0}^{3}\left( 1-e_{r0}^{2}\right) ^{2}\mathfrak{l}_{r0}} 
\notag \\
&&\times \left( \cos \kappa _{1}\cos \kappa _{2}\bar{E}^{SS}\right.  \notag
\\
&&\left. +\sin \kappa _{1}\sin \kappa _{2}\bar{F}^{SS}\right) ~,
\end{eqnarray}%
\begin{eqnarray}
\bar{e}_{rQM} &=&\frac{\pi \left( e_{r0}^{2}+2\right) \eta }{128\left(
1-e_{r0}^{2}\right) ^{5/2}\mathfrak{l}_{r0}}\sum_{k=1}^{2}\chi _{k}^{2}\nu
^{2k-3}w_{k}  \notag \\
&&\times \left[ -4\left( 10e_{r0}^{3}+609e_{r0}^{2}+260e_{r0}\right. \right.
\notag \\
&&\left. +336\right) \sin ^{2}\kappa _{k}\cos 2\zeta _{k}+8\left(
4e_{r0}^{3}\right.  \notag \\
&&\left. +27e_{r0}^{2}+15e_{r0}+12\right)  \notag \\
&&\left. \times \left( 3\cos 2\kappa _{k}+1\right) \right] ~.
\end{eqnarray}%
Here the coefficients $\bar{E}_{k}$ and $\bar{F}_{k}$ of $\bar{e}_{r}$ are
shown in Table \ref{tableeraver}.

\section{Expressing the shape variables at $\protect\chi _{p}=0$ in terms of
the averaged quantities $\mathfrak{\bar{l}}$ and $\bar{e}_{r}$\label%
{inverting}}

In this subsection we invert Eqs. (\ref{lraverexpr}) and (\ref{eraverexpr})
in terms of $\mathfrak{\bar{l}}$ and $\bar{e}_{r}$. We need two steps, the
first is taking the perturbations to first order. Using Eqs (\ref{lraverexpr}%
), (\ref{lravernewtexpr}) and (\ref{tau0expr}) for $\mathfrak{l}_{r0}$ we
get 
\begin{eqnarray}
\mathfrak{l}_{r0} &=&\mathfrak{\bar{l}}_{r}\frac{\mathfrak{T}}{\mathfrak{T}%
_{0}}-\frac{\mathfrak{1}}{\mathfrak{T}_{0}}\left( \mathfrak{\bar{l}}_{rPN}+%
\mathfrak{\bar{l}}_{rSO}+\mathfrak{\bar{l}}_{rSS}\right.  \notag \\
&&\left. +\mathfrak{\bar{l}}_{rQM}\right) ~.
\end{eqnarray}%
Using Eqs (\ref{eraverexpr}), (\ref{eravernewtexpr}) and (\ref{tau0expr})
for $e_{r0}$ we get%
\begin{eqnarray}
e_{r0} &=&\bar{e}_{r}\frac{\mathfrak{T}}{\mathfrak{T}_{0}}-\frac{\mathfrak{1}%
}{\mathfrak{T}_{0}}\left( \bar{e}_{rPN}+\bar{e}_{rSO}+\bar{e}_{rSS}\right. 
\notag \\
&&\left. +\bar{e}_{rQM}\right) ~,
\end{eqnarray}%
In the perturbation terms we have to insert the leading order terms of $%
\mathfrak{l}_{r0}$ and $e_{r0}$%
\begin{eqnarray}
\mathfrak{l}_{r0} &=&\mathfrak{\bar{l}}_{r}~, \\
e_{r0} &=&\bar{e}_{r}~.
\end{eqnarray}%
The results are as follows:%
\begin{equation}
\mathfrak{l}_{r0PN}=\frac{2\bar{e}_{r}(\bar{e}_{r}+1)(\eta -2)}{\mathfrak{%
\bar{l}}_{r}}~,
\end{equation}%
\begin{eqnarray}
\mathfrak{l}_{r0SO} &=&-\frac{\bar{e}_{r}\left( 1-\bar{e}_{r}^{2}\right)
\eta }{2(\bar{e}_{r}-1)\mathfrak{\bar{l}}_{r}^{2}}  \notag \\
&&\times \sum_{k=1}^{2}\left( 4^{2k-3}+3\right) \chi _{k}\cos \kappa _{k}~,
\end{eqnarray}%
\begin{eqnarray}
\mathfrak{l}_{r0SS} &=&-\frac{\chi _{1}\chi _{2}\eta }{32(1-\bar{e}_{r})\bar{%
e}_{r}^{2}\mathfrak{\bar{l}}_{r}^{3}}\left( L_{0}^{SS}\cos \kappa _{1}\cos
\kappa _{2}\right.  \notag \\
&&\left. -K_{0}^{SS}\sin \kappa _{1}\sin \kappa _{2}\right) ~,
\end{eqnarray}%
\begin{eqnarray}
\mathfrak{l}_{r0QM} &=&\frac{\eta }{512\bar{e}_{r}\left( \bar{e}%
_{r}^{2}-1\right) \mathfrak{\bar{l}}_{r}^{3}}\sum_{k=1}^{2}\chi _{k}^{2}\nu
^{2k-3}w_{k}  \notag \\
&&\times \left[ -4\left( 47\bar{e}_{r}^{5}+1338\bar{e}_{r}^{4}+1446\bar{e}%
_{r}^{3}\right. \right.  \notag \\
&&\left. +3444\bar{e}_{r}^{2}+1264\bar{e}_{r}+960\right)  \notag \\
&&\times \sin ^{2}\kappa _{k}\cos 2\zeta _{k}+16\left( 5\bar{e}_{r}^{5}+33%
\bar{e}_{r}^{4}\right.  \notag \\
&&\left. +61\bar{e}_{r}^{3}+84\bar{e}_{r}^{2}+42\bar{e}_{r}+12\right)  \notag
\\
&&\left. \times \left( 3\cos 2\kappa _{k}+\allowbreak 1\right) \right] ~,
\end{eqnarray}%
where the coefficients $L_{0}$ and $K_{0}$ can be found in Table \ref%
{lr0kifcoeffs}. For $e_{r0}$ we get%
\begin{eqnarray*}
e_{r0PN} &=&\frac{1}{\bar{e}_{r}\mathfrak{\bar{l}}_{r}^{2}}\left\{ -\left( 1-%
\bar{e}_{r}^{2}\right) ^{3/2}\left( 3\eta -5\right) \right. \\
&&+\frac{\left( 1+\bar{e}_{r}\right) ^{2}}{2}\left[ \bar{e}_{r}^{2}\left(
11\eta -14\right) \right. \\
&&\left. \left. -2\bar{e}_{r}\left( 5\eta -7\right) +6\eta -10\right]
\right\} ~,
\end{eqnarray*}%
\begin{eqnarray*}
e_{r0SO} &=&-\frac{(\bar{e}_{r}+1)\left( 1-\bar{e}_{r}^{2}\right) \eta }{2%
\mathfrak{\bar{l}}_{r}^{3}} \\
&&\times \sum_{k=1}^{2}\left( 4^{2k-3}+3\right) \chi _{k}\cos \kappa _{k}~,
\end{eqnarray*}%
\begin{eqnarray}
e_{r0SS} &=&-\frac{3\chi _{1}\chi _{2}\eta }{32(1-\bar{e}_{r})\bar{e}_{r}^{3}%
\mathfrak{\bar{l}}_{r}^{4}}\left( E_{0}^{SS}\cos \kappa _{1}\cos \kappa
_{2}\right.  \notag \\
&&\left. -F_{0}^{SS}\sin \kappa _{1}\sin \kappa _{2}\right) ~,
\end{eqnarray}%
\begin{eqnarray}
e_{r0QM} &=&\frac{\eta }{256\left( \bar{e}_{r}^{2}-1\right) \mathfrak{\bar{l}%
}_{r}^{4}}\sum_{k=1}^{2}\chi _{k}^{2}\nu ^{2k-3}w_{k}  \notag \\
&&\times \left[ -4\left( 10\bar{e}_{r}^{5}+753\bar{e}_{r}^{4}+712\bar{e}%
_{r}^{3}\right. \right.  \notag \\
&&\left. +1986\bar{e}_{r}^{2}+664\bar{e}_{r}+672\right)  \notag \\
&&\times \sin ^{2}\kappa _{k}\cos 2\zeta _{k}+8\left( 4\bar{e}_{r}^{5}+39%
\bar{e}_{r}^{4}\right.  \notag \\
&&+59\bar{e}_{r}^{3}+102\bar{e}_{r}^{2}+42\bar{e}_{r}  \notag \\
&&\left. \left. +24\right) \left( 3\cos 2\kappa _{k}+\allowbreak 1\right) 
\right] ~.
\end{eqnarray}%
where the coefficients $L_{0}$ and $K_{0}$ can be found in Table \ref%
{er0kifcoeffs}.

The second step is to derive the 2PN terms. For this we have to use the same
equations as in the first step: Eqs (\ref{lraverexpr}), (\ref{lravernewtexpr}%
) and (\ref{tau0expr}) for $\mathfrak{l}_{r0}$ and Eqs (\ref{eraverexpr}), (%
\ref{eravernewtexpr}) and (\ref{tau0expr}) for $e_{r0}$ 
\begin{eqnarray}
\mathfrak{l}_{r0} &=&\mathfrak{\bar{l}}_{r}\frac{\mathfrak{T}}{\mathfrak{T}%
_{0}}-\frac{\mathfrak{1}}{\mathfrak{T}_{0}}\left( \mathfrak{\bar{l}}_{rPN}+%
\mathfrak{\bar{l}}_{r2PN}\right) ~, \\
e_{r0} &=&\bar{e}_{r}\frac{\mathfrak{T}}{\mathfrak{T}_{0}}-\frac{\mathfrak{1}%
}{\mathfrak{T}_{0}}\left( \bar{e}_{rPN}+\bar{e}_{r2PN}\right) ~.
\end{eqnarray}%
This time, in order to get the 2PN terms we need the previously calculated
1PN expressions of $\mathfrak{l}_{r0}$ and $e_{r0}$. After Taylor-expanding
to 2PN order we get .%
\begin{eqnarray}
\mathfrak{l}_{r02PN} &=&-\frac{(1-\bar{e}_{r}^{2})^{3/2}}{24\bar{e}_{r}^{2}%
\mathfrak{\bar{l}}_{r}^{3}}\sum_{k=0}^{2}L_{0,k}^{2PN}\bar{e}_{r}^{k}  \notag
\\
&&+\frac{(\bar{e}_{r}+1)^{2}}{12\bar{e}_{r}^{2}\mathfrak{\bar{l}}_{r}^{3}}%
\sum_{k=0}^{4}K_{0,k}^{2PN}\bar{e}_{r}^{k}~,
\end{eqnarray}%
\begin{eqnarray}
e_{r02PN} &=&-\frac{\left( 1-\bar{e}_{r}^{2}\right) ^{3/2}}{960\bar{e}%
_{r}^{3}\mathfrak{\bar{l}}_{r}^{4}}\sum_{k=0}^{4}E_{0,k}^{2PN}\bar{e}_{r}^{k}
\notag \\
&&+\frac{(1+\bar{e}_{r})^{2}}{120\bar{e}_{r}^{3}\mathfrak{\bar{l}}_{r}^{4}}%
\sum_{k=0}^{6}F_{0,k}^{2PN}\bar{e}_{r}^{k}~.
\end{eqnarray}%
The coefficients $L_{0}^{2PN}$ and $K_{0}^{2PN}$,are given in Table \ref%
{lr0kifcoeffs}, while the terms $E_{0}^{2PN}$ and $F_{0}^{2PN}$ can be found
in Table \ref{er0kifcoeffs}.

\section{Secular precession angular velocities$\label{appendix2}$}

The averaged precession angular velocities are calculated from Eqs (31-33)
of Ref. \cite{chameleon}\footnote{%
In the equations (B34) of Ref. \cite{Inspiral2} the SS and QM terms have
typos: the $1/2$ factors should be removed. We thank Krisztina K\"{o}v\'{e}r
for pointing this out to us. Due to this, Eqs. (31), (32) and the second
term of Eq. (33) of Ref. \cite{chameleon} contain unnecessary $1/2$ factors
on the rhs (but otherwise all conclusions remain unchanged). In the present
paper these have been corrected and both the instantaneous and secular
dynamics are represented correctly.}.%
\begin{eqnarray}
\overline{\mathbf{\omega }_{\mathbf{i}}\cdot \mathbf{\hat{A}}_{\mathbf{N}}}
&=&\frac{2\eta \pi }{\mathfrak{T}~\mathfrak{\bar{l}}_{r}^{3}}\left( \nu
^{2j-3}\chi _{j}\sin \kappa _{j}\cos \zeta _{j}\right.  \notag \\
&&\left. +3w_{i}\chi _{i}\sin \kappa _{i}\cos \zeta _{i}\right)
\end{eqnarray}%
\begin{eqnarray}
\overline{\mathbf{\omega }_{\mathbf{i}}\cdot \mathbf{\hat{Q}}_{\mathbf{N}}}
&=&\frac{2\eta \pi }{\mathfrak{T}~\mathfrak{\bar{l}}_{r}^{3}}\left( \nu
^{2j-3}\chi _{j}\sin \kappa _{j}\sin \zeta _{j}\right.  \notag \\
&&\left. +3w_{i}\chi _{i}\sin \kappa _{i}\sin \zeta _{i}\right)
\end{eqnarray}%
\begin{eqnarray}
\overline{\mathbf{\omega }_{\mathbf{i}}\cdot \mathbf{\hat{L}}_{\mathbf{N}}}
&=&\frac{\eta \pi }{\mathfrak{T}~\mathfrak{\bar{l}}_{r}^{3}}\left[ ~%
\mathfrak{\bar{l}}_{r}\left( 4+3\nu ^{3-2i}\right) \right.  \notag \\
&&\left. +2\nu ^{2j-3}\chi _{j}\cos \kappa _{j}\right]
\end{eqnarray}

It is easy to see by checking the leading order term of $\mathfrak{T}$ that
as $\bar{e}_{r}$ goes to $1$ the precession becomes increasingly small. It
is explained by the fact that on parabolic orbits, when $e_{r}=1$, the
motion becomes unbound, and there is no well defined period, thus no
precession.

\end{document}